\documentclass[12pt]{article}
\usepackage[xdvi]{graphicx}
\usepackage{amssymb}
\newcommand{\bea}{\begin{eqnarray}}
\newcommand{\eea}{\end{eqnarray}}

\makeatletter
\@addtoreset{equation}{section}
\makeatother

\newcommand{\simg}{%
\hspace{0.3em}\raisebox{0.4ex}{$>$}\hspace{-0.75em}\raisebox{-.7ex}{$\sim$}\hspace{0.3em}} 

\newcommand{\siml}{%
\hspace{0.3em}\raisebox{0.4ex}{$<$}\hspace{-0.75em}\raisebox{-.7ex}{$\sim$}\hspace{0.3em}}

\begin{document}
\begin{titlepage}
\begin{flushright}
OU-HET 639/2009
\end{flushright}

\vspace{25ex}

\begin{center}
{\Large\bf 
Two-loop corrections and predictability on orbifold
}
\end{center}

\vspace{1ex}

\begin{center}
{\large
Nobuhiro Uekusa
}
\end{center}
\begin{center}
{\it Department of Physics, 
Osaka University \\
Toyonaka, Osaka 560-0043
Japan} \\
\textit{E-mail}: uekusa@het.phys.sci.osaka-u.ac.jp
\end{center}


\vspace{3ex}

\begin{abstract}

We study quantum loop corrections to
two-point functions and extraction of physical
quantities 
in a five-dimensional 
$\phi^4$ theory on an orbifold.
At two-loop level,
we find that
divergence for quartic derivative terms
of $(p^2)^2$ appear as Lagrangian
terms in the bulk. The counterterms are needed
and these terms make propagators have two poles.
With this effect taken into account,
corrections to masses are derived. 
We show that for extraction of physical quantities
for two-point functions an
ultraviolet cutoff must be orders of
magnitude larger compared to a compactification
scale.
Even higher derivative corrections
at higher loop levels are also discussed.

\end{abstract}
\end{titlepage}

\section{Introduction}

The description of physical quantities has been
widely developed in theory based on the action
principle.
The strength and the form of interactions affect
physical quantities involving
virtual processes by quantum loop corrections.
It is a key ingredient how interactions are 
taken into account in the action integral.
As a principle,
theory has invariance
such as Poincar\'{e} invariance.
Because invariances of a theory lead to
conserved currents,
they are significant constraints.

Renormalizability is
another important element of
a restriction to theory.
Once a finite number of all the renormalizable 
interactions are written down,
new counterterms are not required.
This makes perturbation valid.
On the other hand, invariance of theory such as
Poincar\'{e} invariance does not have a crucial
role to forbid
non-renormalizable interactions.
If a non-renormalizable interaction is added,
new counterterms would be needed.
The new terms would give rise to the corresponding
contributions
and the number of counterterms can be infinite.
Renormalizability seems to be additional requirement
in perturbation.
In extracting physical quantities, however,
renormalizability is not compulsory
from the point of view that
non-renormalizable interactions are irrelevant operators and that the contributions to
physical quantities are negligible.
In four dimensions, usually non-renormalizable
interactions are supposed to be suppressed by an 
ultraviolet momentum cutoff
of a theory.
Effective theory with a large cutoff can be
predictable without requiring renormalizability.

It is nontrivial whether only four-dimensional
theory is privileged to yield physical accuracy
in a non-renormalizable theory.
A characteristic property in four dimensions 
is that renormalizable and non-renormalizable
interactions coexist.
In a theory with compactified extra dimensions,
fields as four-dimensional modes can have dimension-four
operators which are similar to
renormalizable terms in four dimensions.
If coefficients of other operators are small,
such a theory might be predictable with a
certain accuracy.
The coefficients of higher dimensional
operators are unknown and
should be eventually determined.
For addressing this problem, there are some
attitudes.
One is to try to construct a consistent theory 
to specify
all the non-renormalizable interactions.
Another is to search for rules or orders for
possible interactions at each given loop level
in a non-renormalizable theory.
In this paper, our standing point is to aim for
identifying contributions
of higher-dimensional operators
appearing at two-loop level.

Since new frameworks for solving the hierarchy 
problem have been proposed~%
\cite{ArkaniHamed:1998rs,rs},
quantum loop corrections of interactions
in a bulk 
have been studied in various models
with compactified extra dimensions.
As a radical possibility,
all the fields of the standard model might 
propagate in a bulk~\cite{Appelquist:2000nn}.
If gauge bosons propagate in a bulk,
there is a mechanism of dynamical gauge 
symmetry breaking~\cite{hm}.
One of remarkable features in orbifold models
is that loop effects of bulk fields
produce infinite contributions 
to require renormalization by 
four-dimensional couplings on boundaries~%
\cite{Georgi:2000ks}.
Boundary terms can 
be mass and kinetic energy terms
and higher derivative operators can be needed 
as counterterms for loop corrections~%
\cite{vonGersdorff:2002as}-%
\cite{Ghilencea:2004sq}.

Recently, 
higher derivative terms have been utilized
for
a phenomenologically accessible idea in the
Lee-Wick standard model~%
\cite{Grinstein:2007mp}.
Here the higher derivative propagator contains
two poles and a cancellation 
between contributions from the two poles removes the
quadratic
divergence associated with the Higgs mass.
In the context of large extra dimensions,
it has been shown that
a new type of 
divergence which corresponds to a higher derivative
operator is generated by
quantum corrections from the exchange of virtual 
Kaluza-Klein gravitons~\cite{Wu:2008rr}.
It gives 
an explanation for solving the hierarchy problem
as an ultraviolet cutoff of the theory 
is the TeV scale.
Aspects such as subtlety about
massive ghost fields have also been
discussed in this model~\cite{Rodigast:2009en}.
The scenario of the Lee-Wick standard model
is interesting but
the effect of higher derivative operators  
seems comparable to that of dimension-four 
operators.
In this paper, we pursue 
a conventional approach based on such a view
that irrelevant operators are small
in an effective theory
while we treat poles of
propagators in a way developed 
in the Lee-Wick standard models.

We calculate quantum loop corrections to
two-point functions  
in a five-dimensional $\phi^4$ theory 
on an orbifold $S^1/Z_2$.
As in four dimensions, divergences
for mass terms and wave fucntions
are found at one-loop and two-loop levels,
respectively
At two-loop level,
we find that divergence for quartic derivative terms 
of $(p^2)^2$ appears for the bulk.
The counterterms are needed and
these terms make
propagators have two poles.
With this effect taken into account,
corrections to two-point functions are derived. 
The contributions of the quartic derivative
terms to masses depend on both of an ultraviolet
cutoff denoted as $\Lambda$ and
a size of an extra dimension denoted as $L$.
We show that the contributions can be
extracted with a moderate fine-tuning for 
$\Lambda L \simg 10^2$ and 
that they cannot be fixed for
$\Lambda L \sim 10$.
Therefore,
for extraction of physical quantities
for two-point functions an
ultraviolet cutoff must be orders of
magnitude larger compared to a compactification
scale.
This behavior is in agreement
with the conventional observation
that contributions of higher-dimensional
operators are small for a large cutoff.

We also discuss corrections with multiple poles
at higher loop levels.
An example of higher derivative terms beyond $(p^2)^2$ is 
given at four-loop level.
It is found that a similar large cutoff tends to be needed.

The paper is organized as follows. 
In Sec.~\ref{sec:model}, we present
our model including fields and action integrals.
In Sec.~\ref{sec:oneloop},
two-point functions at tree and one-loop levels are
given.
In Sec.~\ref{sec:twoloop}, we calculate
quantum corrections
for two-point functions at two-loop level.
It is found that divergence of $(p^2)^2$ terms appears.
In Sec.~\ref{sec:counter},
by starting with the action integral including
$(p^2)^2$ terms, we derive corrections
for two-point functions.
Here dependence of the corrections on $\Lambda L$
is given.
In Sec.~\ref{sec:high},
higher-loop corrections are discussed.
An exemplification is given at four-loop level.
We conclude in Sec.~\ref{sec:concl}
with some remarks.
Details of calculations and
formulas are shown in appendices.

\section{Model \label{sec:model}}

We start with the action for the real scalar field
$\phi(x,y)$,
\bea
  S&\!\!\!=\!\!\!&
    \int d^4x \int_0^L dy
     \left({1\over 2}(\partial_\mu \phi)^2
     -{1\over 2}(\partial_y \phi)^2
     -{1\over 2}m^2 \phi^2
     -\lambda_5 \phi^4\right) .
     \label{action}
\eea
Greek indices $\mu$ run over 0,1,2,3
and fifth index is denoted as $y$.
We use a metric $\eta_{\mu\nu}=\textrm{diag}(1,-1,-1,-1)$.
The extra-dimensional space is compactified on $S^1/Z_2$,
where the fundamental region is $0\leq y \leq L$.
The five-dimensional spacetime is flat.
The boundary conditions for the scalar $\phi(x,y)$ are
\bea
   \phi(x, -y) = \phi(x, y) ,\qquad
   \phi(x, L-y) = \phi(x, L+y) .
\eea   
The mode expansion is given by
\bea
   \phi (x,y) &\!\!\!=\!\!\!&
 \sqrt{1\over L} \, \phi_0(x) 
 + \sum_{n=1}^\infty
  \sqrt{2\over L} \, \cos\left(m_n y\right)
    \phi_n (x) ,
    \label{sme}
\eea
where the mass for $\phi_n(x)$ is $m_n={n\pi\over L}$.
In calculating loop corrections,
it is convenient to employ
$-\infty \leq n\leq \infty$ for a sum with respect to 
$n$. 
Eq.~(\ref{sme}) is rewritten as
\bea    
  \phi (x,y) &\!\!\!=\!\!\!&  \sqrt{1\over 2L}
    \left(
     (\sqrt{2}-1)
       \phi_0(x) 
    +\sum_{n=-\infty}^\infty
    \cos\left(m_n y\right)
      \phi_{|n|}(x) 
   \right) .
\eea
Substituting this equation into the action
(\ref{action}) leads to
\bea
  S &\!\!\!=\!\!\!&
   \int d^4 x \left\{
    {1\over 2} (\partial_\mu \phi_0)^2
    -{1\over 2}m^2 \phi_0^2
  +\sum_{n=1}^\infty
    \left({1\over 2} (\partial_\mu \phi_{n})^2
  -{1\over 2} (m^2 + m_n^2 )\phi_{n}^2 \right) 
  \right.
\nonumber
\\
  && -{\lambda_5 \over 4L}
    \left(
      \left( (\sqrt{2}-1)^4 +4(\sqrt{2}-1)^3
   \right) \phi_0^4 
     \right.
  +6(\sqrt{2}-1)^2
   \sum_{n=-\infty}^\infty
     \phi_0^2 \phi_{|n|}^2
\nonumber
\\
  && + 4(\sqrt{2}-1) \sum_{n=-\infty}^\infty
  \sum_{\ell=-\infty}^\infty
    \phi_0 \phi_{|n|} \phi_{|\ell|}
      \phi_{|n+\ell|}
  \left.\left.
 +\sum_{n=-\infty}^\infty
   \sum_{\ell=-\infty}^\infty
   \sum_{s=-\infty}^\infty
     \phi_{|n|} \phi_{|\ell|}
     \phi_{|s|} \phi_{|n+\ell+s|} \right) 
 \right\} ,
  \label{actionk}
\eea
where the kinetic energy and mass terms
are denoted as the sum of zero mode
and modes $1\leq n\leq \infty$
so that the modes of
quadratic terms correspond to degrees of 
freedom of propagation.
Here $\lambda\equiv \lambda_5/L$ is dimensionless.
Infinite sum with respect to modes 
is the effect of the 
five-dimensional origin.
Even if there are only dimension-two
and dimension-four operators,
the five-dimensional effect gives rise to
new divergent terms, as we will see.
The interaction terms have invariance under 
$n\leftrightarrow -n$.
The quadratic terms of scalars are
diagonal with respect to modes $n$.

At boundaries, $\phi^2$ are expanded as
\bea
    \left.\phi^2 \right|_{y=0}
     &\!\!\!=\!\!\!& {1\over 2L}
       \left((\sqrt{2}-1)^2\phi_0^2
    +2(\sqrt{2}-1)\sum_{n=-\infty}^\infty
       \phi_0\phi_{|n|}
  +\sum_{n=-\infty}^\infty \sum_{\ell=-\infty}^\infty
       \phi_{|n|}\phi_{|\ell|}\right) ,
\\ 
   \left.\phi^2 \right|_{y=L}
     &\!\!\!=\!\!\!& {1\over 2L}
       \left((\sqrt{2}-1)^2\phi_0^2
    +2(\sqrt{2}-1)\sum_{n=-\infty}^\infty
       (-1)^n\phi_0\phi_{|n|}
     \right.
\nonumber
\\
  && \left.
  +\sum_{n=-\infty}^\infty \sum_{\ell=-\infty}^\infty
       (-1)^{n+\ell}\phi_{|n|}\phi_{|\ell|}\right) .
\eea
From this equation, quadratic terms at boundaries are
\bea
  && \int_0^L dy \, \phi^2
    \left( \delta(y) +\delta(y-L) \right) 
\nonumber
\\
 &\!\!\!=\!\!\!&
   {1\over L}
     \left( (\sqrt{2}-1)^2 \phi_0^2 
       +2(\sqrt{2}-1)\sum_{n=-\infty}^\infty
         \phi_0 \phi_{|2n|}
     +\sum_{n=-\infty}^\infty
      \sum_{s=-\infty}^\infty
      \phi_{|n|} \phi_{|n+2s|} \right),
  \label{ff2n}
\eea
which have non-diagonal components
with respect to modes $n$.

By a dimensional analysis with
$[\phi]=[\textrm{mass}]^{3\over 2}$
and $[\lambda_5]=[\textrm{mass}]^{-1}$,
possible Lagrangian counterterms are expected 
in terms of power of $\lambda_5$ as
\bea    
   (\partial \phi)^2 , ~~
        \lambda_5 \phi^4 ,~~
    \lambda_5^2 (\partial^2 \phi)^2 , ~~ 
  \lambda_5^3 (\partial \phi)^2 \phi^2 , ~~
  \lambda_5^4 \phi^6 , ~~
  \lambda_5^4 (\partial^3 \phi)^2 , ~~ 
   \lambda_5^6 (\partial \phi)^2 \phi^4 , ~~
  \cdots ,
  \label{diman}
\eea
where vector indices are contracted.

\section{Two-point functions at tree and one-loop levels
\label{sec:oneloop}}

At tree level, the two-point functions for 
$\phi_0(x)\phi_0(x)$ and $\phi_n(x)\phi_n(x)$ are
\bea
   D(x-w) &\!\!\!=\!\!\!&
  \int {d^4p\over (2\pi)^4} \,
   {i\over p^2 -m^2 +i\epsilon} \,
  e^{-ip\cdot (x-w)} ,
\\
    D_n(x-w) &\!\!\!=\!\!\!&
  \int {d^4p\over (2\pi)^4} \,
   {i\over p^2 -m^2 -m_n^2 +i\epsilon} \,
  e^{-ip\cdot (x-w)} ,    
   \label{dnprop}
\eea
respectively. We will omit $i\epsilon$ hereafter.

\begin{figure}[htb]
\begin{center}
  \includegraphics[width=2cm]{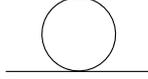}
 \caption{One-loop diagram for two-point functions.
 \label{fig:1loop}}
\end{center}
\end{figure}

The diagram for one-loop two-point functions 
is shown in Fig.~\ref{fig:1loop}.
Here external lines can take 
$\phi_0$ or $\phi_{|n|}$.
One-loop amplitudes are calculated from
\bea
  \left\langle
    \phi_0 (x) \phi_0 (w) \,
      i\int d^4 x \,{\cal L} 
    \right\rangle ,
  \quad
    \left\langle
    \phi_0 (x) \phi_{|f|}(w) \,
      i\int d^4 x \,{\cal L} 
    \right\rangle ,
 \quad
    \left\langle
    \phi_{|f|} (x) \phi_{|g|} (w) \,
      i\int d^4 x \,{\cal L} 
    \right\rangle .
\eea
We obtain the one-loop contribution 
for the two-point function with
the external lines $\phi_{|f|}\phi_{|g|}$ as
\bea
 \tilde{G}_{|f||g|}&\!\!\!=\!\!\!&
  - 3i\lambda \bigg( 
   (\sqrt{2}-1)^2
  \delta_{f0} \delta_{g0}
  {\cal D}     
 + \delta_{f0} \delta_{g0}
  \sum_{n=-\infty}^\infty
    {\cal D}_n
    + \, \delta_{|f||g|}
  \sum_{n=-\infty}^\infty
    {\cal D}_n
\nonumber
\\
  && +(\sqrt{2}-1)  \, 
  \sum_{n=-\infty}^\infty 
  \left(\delta_{f0} \delta_{|g||2n|}
   + \delta_{g0}\delta_{|f||2n|}\right)
   {\cal D}_n       
   + 
  \sum_{n=-\infty}^\infty \delta_{|f||g+2n|}
    {\cal D}_n
  \bigg) .
   \label{eq1loopt}
\eea
Here the two momentum integrals are given as
\bea
  {\cal D} =
\int {d^4 k\over (2\pi)^4}
     \, {i\over k^2 -m^2} ,
\qquad
  {\cal D}_n =
\int {d^4 k\over (2\pi)^4}
     \, {i\over k^2 -m^2 -m_n^2} .
\eea
The twp-point functions with the external lines
$\phi_0\phi_0$ and $\phi_0\phi_{|f|}$ can be
derived as limits of $\phi_{|f|}\phi_{|g|}$.
The details of calculations for derivation
of Eq.~(\ref{eq1loopt}) are given in 
App.~\ref{ap:c1}.
A note given in contraction with respect to $n$
is that
$\sum_{n=-\infty}^\infty \delta_{|\ell| |n|}
    \delta_{|s| |n|}
     \neq \delta_{|\ell| |s|}$. 
Here
\bea
   \sum_{n=-\infty}^\infty \delta_{|\ell| |n|}
   \delta_{|s| |n|}
  = 2(\delta_{\ell s} + \delta_{\ell , -s})
    -3 \delta_{\ell 0} \delta_{s0} ,
\qquad
  \delta_{|\ell| |s|}
  =
  \delta_{\ell s} + \delta_{\ell , -s}
  -\delta_{\ell 0} \delta_{s0} .
\eea

Now we evaluate 
the one-loop contribution (\ref{eq1loopt})
as Lagrangian terms of interactions.
Multiplying $\tilde{G}_{|f||g|}$ by
$\phi_{|f|}\phi_{|g|}$ and summing with respect to
$f,g$ lead to
\bea
  &&
 \sum_{f=-\infty}^\infty 
  \sum_{g=-\infty}^\infty 
    \phi_{|f|}\phi_{|g|} \tilde{G}_{|f||g|}
 =
    -3i\lambda\left(
   (\sqrt{2}-1)^2 \phi_0^2 {\cal D}
  +2(\sqrt{2}-1) 
   \sum_{n=-\infty}^\infty \phi_0 \phi_{|2n|} {\cal D}_n
  \right.
\nonumber
\\
  && \qquad\qquad
  +\sum_{f=-\infty}^\infty \sum_{n=-\infty}^\infty
    \phi_{|f|} \phi_{|f+2n|} {\cal D}_n
   \left.
   +\left( \phi_0^2 + \sum_{\ell=-\infty}^\infty
  \phi_{|\ell|}^2 \right)
  \sum_{n=-\infty}^\infty {\cal D}_n \right) .
    \label{eval}
\eea
This one-loop corrections include bulk and boundary
divergences.

A bulk mass term is expanded as
\bea
   - \int_0^L dy \, m^2 \phi^2
  =-{1\over 2} m^2 \left(
   \phi_0^2
     +\sum_{\ell=-\infty}^\infty \phi_{|\ell|}^2
      \right) .
\eea
From Eq.~(\ref{eval}),
bulk divergent terms at one-loop level
are
\bea
   \left(\phi_0^2 +\sum_{\ell =-\infty}^\infty
     \phi_{|\ell|}^2\right)
       \sum_{n=-\infty}^\infty {\cal D}_n .
     \label{bulkterm}
\eea
The divergent part of the integral is
\bea
 \left.
 \sum_{n=-\infty}^\infty {\cal D}_n  \right|_{%
 \textrm{\scriptsize div}}
 = 
{1\over 2^4 \pi^{5\over 2}}
     \left({2\over 3}\Lambda^2 
 -2m^2\right) \Lambda L . 
  \label{1bulk}
\eea
The details of calculations are given
in App.~\ref{ap:dint}.
From Eqs.~(\ref{bulkterm}) and (\ref{1bulk}),
the bulk divergent term at one-loop level
is obtained as
\bea
   {1\over 2^2 \pi^{5\over 2}}
    \int_0^L dy \,
     \left({1\over 3}\Lambda^2 
 -m^2\right) \Lambda L \phi^2.
\eea     
The divergence of $\Lambda^3$ is expected from
a five-dimensional integral of a propagator,
\bea
   \sum_{n=-\infty}^\infty
     \int {d^4 k\over (2\pi)^4}
     {i\over k^2 -m^2 -m_n^2}
   \to
   \int {d^5 \bar{k} \over (2\pi)^5}
     {i\over \bar{k}^2 -m^2}
   \sim \Lambda^3 ,
    \label{45to}
\eea
where $\bar{k}= (k,k_y)$.

In Eq.~(\ref{eval}), 
the other divergent terms have contractions
between external mode indices
and internal mode indices.
The sum with respect to modes is not 
independent of external lines.
It is regarded as the structure that
each mode has four-dimensional divergence,
\bea
  \int {d^4k \over (2\pi)^4}
   \, {i\over k^2 -m_n^2 -M^2} 
 &\!\!\!=\!\!\!&
  {1\over 16\pi^2}
    \int_0^{\Lambda^2}
    d(k_E^2) \,
     {k_E^2\over k_E^2 +m_n^2 +M^2} 
\nonumber
\\
  &\!\!\!=\!\!\!&
   {1\over 16\pi^2}
   \left(
   \Lambda^2 
   -(m_n^2 +M^2)
    \log {\Lambda^2\over m_n^2 +M^2} \right) ,
    \label{1cutoffint}
\eea
where $M=m$ for ${\cal D}_n$.

From Eq.~(\ref{ff2n}),
boundary terms including $\partial_y$ are
\bea
  &&  \int_0^L dy \, {1\over 4}  \left(
      \partial_y^2 \phi\cdot \phi
     +\phi\partial_y^2 \phi
   -2 \partial_y \phi\partial_y \phi\right)
   \left( \delta(y) +\delta(y-L) \right) 
\nonumber
\\
  &\!\!\!=\!\!\!&
  - {1\over 4L}
    \left(\sum_f \sum_n
   \left( m_f 
    -m_{f+2n} 
     \right)^2 \phi_{|f|}
     \phi_{|f+2n|}
   +2(\sqrt{2}-1) \sum_n \left(m_{2n}^2\right)
    \phi_0 \phi_{|2n|}
    \right) 
\nonumber
\\
  &\!\!\!=\!\!\!&
  - {1\over L}
    \left(\sum_f \sum_n
   m_n ^2 \phi_{|f|}
     \phi_{|f+2n|}
   +2(\sqrt{2}-1) \sum_n m_n^2
    \phi_0 \phi_{|2n|}
    \right)  .
\eea
From this equation and Eq.~(\ref{ff2n}), the boundary terms in one-loop two-point function corresponds to
\bea
  &&  \int_0^L dy \, 
   \left\{-m^2 \phi^2 + {1\over 4}  \left(
      \partial_y^2 \phi\cdot \phi
     +\phi\partial_y^2 \phi
   -2 \partial_y \phi\partial_y \phi\right)\right\}
   \left( \delta(y) +\delta(y-L) \right)
\nonumber
\\
  &\!\!\!=\!\!\!&
  -{1\over L}
    \left(
   (\sqrt{2}-1)^2 m^2 \phi_0^2
 +2(\sqrt{2}-1) \sum_{n=-\infty}^\infty
   \left(m^2+ m_n^2\right)
    \phi_0 \phi_{|2n|}
  \right.
\nonumber
\\
  && \left.
 + \sum_{f=-\infty}^\infty \sum_{n=-\infty}^\infty
   \left(m^2 + m_n ^2\right) \phi_{|f|}
     \phi_{|f+2n|}   
    \right)  .
\eea
On the boundaries, as in a four-dimensional theory
there is quadratic divergence for mass terms.

\section{Two-point functions at two-loop level
\label{sec:twoloop}}

\begin{figure}[htb]
\begin{center}
 \includegraphics[width=2cm]{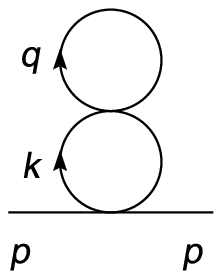}  \qquad\quad
 \includegraphics[width=4cm]{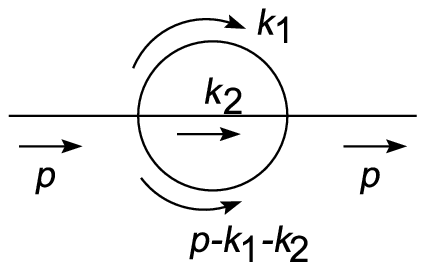} \\
\hspace{-2.5em} (a) \qquad\qquad\qquad\qquad\quad (b)
 \caption{Two-loop diagrams for two-point functions
 \label{fig:twoloop}}
\end{center}
\end{figure}

At two-loop level,
there are two one-particle irreducible diagrams
shown in Fig.~\ref{fig:twoloop}.
In Fig.~\ref{fig:twoloop}(a),
the internal loops are independent of the external momentum
$p$.
This means that the diagram does not contribute to
wave functions and higher derivative terms.
At one-loop level, mass terms have been 
introduced for both of 
bulk and boundary Lagrangian terms.
We will not discuss this diagram further.

The two-loop diagram we calculate is
the diagram drawn in Fig.~\ref{fig:twoloop}(b).
The amplitude is calculated from
\bea
  &&\left\langle \phi\phi
    \left(-{1\over 2}\right)
      \int d^4 x_1 {\cal L} \cdot
      \int d^4 x_2 {\cal L} \right\rangle .
      \label{generall}
\eea
The last term in the action (\ref{actionk})
yield the contribution to the diagram in 
Fig.~\ref{fig:twoloop}(b),
\bea
  &&
        \left\langle \phi_{|f|}(x)\phi_{|g|}(y)
    \left(-{1\over 2}\right)
      \left(-{\lambda\over 4}\right)^2
      \int d^4 x_1  \phi_{|n|} \phi_{|\ell|}
       \phi_{|s|}\phi_{|n+\ell+ s|}
      \int d^4 x_2 
   \phi_{|a|}\phi_{|b|}
    \phi_{|c|} \phi_{|a+b+c|}
  \right\rangle
\nonumber
\\
  &\!\!\!\stackrel{\textrm{\scriptsize Fig.}}{=} \!\!\!&
  -6\lambda^2 
  \int d^4x_1 d^4x_2 \,
     \delta_{|f||n|} \,
     \delta_{|g||a|} \,
     \delta_{|\ell||b|} \,
     \delta_{|s||c|} \,
     \delta_{|n+\ell+s||a+b+c|} \,
  D_f D_g D_\ell D_s D_{n+\ell+s}
\nonumber
\\
  &\!\!\!\equiv \!\!\!& \tilde{G}_{|f||g|} (p)
  \tilde{D}_f (p) \tilde{D}_g (p)
 , \label{44}
\eea
where the propagator $D_f$ is given in 
Eq.~(\ref{dnprop}) and its Fourier transformation 
is given by
$\tilde{D}_f (p)= i/(p^2 -m^2 -m_f^2)$.
Expansion of products of
Kronecker $\delta$ is shown in Appendix~\ref{ap:d}.
The corresponding Lagrangian terms are
obtained as
\bea
&& \sum_{f=-\infty}^\infty
 \sum_{g=-\infty}^\infty
   \phi_{|f|} \phi_{|g|}  \tilde{G}_{|f||g|}
\nonumber
\\
  &\!\!\!=\!\!\!&
  -6 \lambda^2
  \sum_{\ell=-\infty}^\infty 
  \sum_{s=-\infty}^\infty
   \int {d^4 k_1\over (2\pi)^4}
  \int {d^4 k_2 \over (2\pi)^4}
 \, \tilde{D}_\ell (k_1) \tilde{D}_s(k_2)
\nonumber
\\
 && \times \left[
  \sum_{f=-\infty}^\infty
   8\left(
     \tilde{D}_{f+\ell +s} \phi_{|f|} \phi_{|f|}
     +\tilde{D}_{f+\ell +s} 
  \phi_{|f|} \phi_{|f+2(\ell+s)|}
     +2\tilde{D}_{f+\ell +s}\phi_{|f|} \phi_{|f+2s|}
     \right)
  \right.
\nonumber
\\
&& -24 \tilde{D}_0 \phi_{|\ell+s|} \phi_{|\ell+s|}
 -24\tilde{D}_0 \phi_{|\ell+s|} \phi_{|s-\ell|}
  +24 \delta_{s0} \tilde{D}_0 \phi_{|\ell|}\phi_{|\ell|}
   -25 \delta_{\ell 0}\delta_{s0}
  \tilde{D}_0\phi_0 \phi_0
\nonumber
\\
 &&
  -6 \tilde{D}_{\ell+s} \phi_0\phi_0 
   -8\tilde{D}_{\ell+s} \phi_{|2(\ell+s)|} \phi_0
  +8\tilde{D}_0\delta_{\ell+s,0}\phi_0 \phi_0 
   -16 \tilde{D}_{\ell-s} \phi_{|2s|}\phi_0 
\nonumber
\\
  && \left.
 +8\tilde{D}_0 \delta_{\ell s}\phi_{|2\ell|}\phi_0
  +16\delta_{s0}\tilde{D}_\ell \phi_0 \phi_0 
  +16\delta_{s0}\tilde{D}_\ell \phi_{|2\ell|}\phi_0 \right] 
 , \label{2loop44}
\eea
where
$\tilde{D}=\tilde{D}(p-k_1-k_2)$  in the brackets.
In this equation, 
the term with three sums 
of $\sum_\ell \sum_s \sum _f$
\bea
  -48\lambda^2 \sum_{\ell=-\infty}^\infty 
  \sum_{s=-\infty}^\infty 
   \int {d^4 k_1\over (2\pi)^4}
  \int {d^4 k_2 \over (2\pi)^4} \,
  \tilde{D}_\ell(k_1)\tilde{D}_s(k_2)
    \sum_{f=-\infty}^\infty 
  \tilde{D}_{f+\ell +s}(p-k_1-k_2)
        \phi_{|f|}\phi_{|f|} ,
      \label{lsf}
\eea
comes only from Eq.~(\ref{44}) among various
terms in Eq.~(\ref{generall}).
Eq.~(\ref{lsf}) is the full term with three sums 
of $\sum_\ell \sum_s \sum _f$, which is
related to divergent $(p^2)^2$ terms.

Now we evaluate the integral
\bea
 && \sum_{\ell=-\infty}^\infty 
  \sum_{s=-\infty}^\infty 
   \int {d^4 k_1\over (2\pi)^4}
  \int {d^4 k_2 \over (2\pi)^4} \,
  \tilde{D}_\ell(k_1)\tilde{D}_s(k_2)
  \tilde{D}_{f+\ell +s}(p-k_1-k_2)
\nonumber
\\
  &\!\!\!=\!\!\!&\sum_{\ell=-\infty}^\infty
    \int {d^4k_1 \over (2\pi)^4}
     \,
   {i\over k_1^2 -m^2 -m_\ell^2}
\nonumber
\\
  && \times
  \left(
    \sum_{s=-\infty}^\infty
      \int{d^4 k_2 \over (2\pi)^4}
      \,
      {i\over k_2^2 -m^2 -m_s^2}
      {i\over (p-k_1 -k_2)^2 
      -m^2 -m_{f+\ell+s}^2}
    \right) ,
    \label{rewr}
\eea
In order to extract dependence on an external momentum
$p$, we expand momentum-dependent part of 
the integrand as
\bea
  {1\over (p-k)^2 -m^2}
  &\!\!\!=\!\!\!&
  {1\over k^2 -m^2}
  +\left[
    -{1\over (k^2 -m^2)^2}
    +{k^2
    \over (k^2 -m^2)^3}\right]
     p^2
\nonumber
\\
  && +
   \left[
   {1\over (k^2 -m^2)^3}
   -{3k^2 \over (k^2 -m^2)^4}
   +{2(k^2)^2\over (k^2 -m^2)^5}
   \right] (p^2)^2
\nonumber
\\
  && +{\cal O}\left(
   {1\over (k^2 -m^2)^4} (p^2)^3\right) ,
\eea
where odd $k$ has been dropped as it vanishes
for $k$ integral.  
It will be seen that 
$(p^2)^s~(s\geq 3)$ terms are convergent.
For simplicity, we proceed the evaluation of
the integral for $m=0$. The case with
nonzero $m$ can be analyzed in a similar way.
The $(p^2)^2$ terms in Eq.~(\ref{rewr})  are
\bea
   &&\sum_{\ell=-\infty}^\infty \sum_{s=-\infty}^\infty
   \int {d^4k_1\over (2\pi)^4}
   \int {d^4k_2 \over (2\pi)^4}
   \, {i\over k_1^2 -m_\ell^2}
   {i\over k_2^2 -m_s^2}
    i(p^2)^2
\nonumber
\\
  &&\times
   \left[
   {1\over ((k_1+k_2)^2 -m_{f+\ell+s}^2)^3}
  -{3(k_1+k_2)^2\over
   ((k_1+k_2)^2 - m_{f+\ell+s}^2)^4}
 +{2((k_1+k_2)^2)^2\over
  ((k_1+k_2)^2 -m_{f+\ell+s}^2)^5} \right] .
\nonumber
\\
  && 
  \label{p4b}
\eea
From this equation,
we find the divergence for $(p^2)^2$ terms,
\bea
    i(p^2)^2 {L^2 \over 105 (4\pi)^4}
    \log(\Lambda^2 L^2) .
    \label{2loopr}
\eea 
For a calculation of the integral,
equations and identities are shown in 
App.~\ref{ap:dint}.

The Lagrangian term for this $(p^2)^2$ divergence is
obtained by substituting Eq.~(\ref{2loopr}) into
Eq.~(\ref{lsf}) as
\bea
 -{i\over 35(2\pi)^4}
 \sum_{f=-\infty}^\infty 
 \log(\Lambda^2 L^2) \cdot \lambda_5^2 
  (p^2)^2 \phi_{|f|}\phi_{|f|} .
  \label{result1}
\eea
The equation (\ref{result1}) has the sum of diagonal
components with respect to mode $f$
so that the contributions are bulk terms.
The divergence of $\log \Lambda$ for $(p^2)^2$ terms
in the bulk is expected from an interpretation
of Eq.~(\ref{p4b}) as
a five-dimensional integral,
\bea
   &&\sum_{\ell=-\infty}^\infty \sum_{s=-\infty}^\infty
   \int {d^4k_1\over (2\pi)^4}
   \int {d^4k_2 \over (2\pi)^4}
   \, {1\over k_1^2 -m_\ell^2}
   {1\over k_2^2 -m_s^2}
   {1\over ((k_1+k_2)^2 -m_{f+\ell+s}^2)^3}
\nonumber
\\
  &\!\!\!\to\!\!\!& 
   \int {d^5 \bar{k}_1 \over (2\pi)^5}
     \int {d^5 \bar{k}_2 \over (2\pi)^5}
     \,
     {1\over \bar{k}_1^2}   
     {1\over \bar{k}_2^2}
     {1\over ((\bar{k}_1+\bar{k}_2)^2)^3}
   \sim \log \Lambda ,
   \label{k3d}
\eea
where $\bar{k}= (k,k_y)$.
In five dimensions,
the emergence of higher derivative terms
as an effective field theory has also been
shown in a method with a
space-dependent cutoff~%
\cite{Lewandowski:2002rf,Lewandowski:2003be,Uekusa:2003nh}.

Boundary terms have contractions
between external mode indices
and internal mode indices.
The sum with respect to modes is not 
independent of external fields $\phi_{|s|}$.
On boundaries, there is no divergence for $(p^2)^2$
because the number of summation is
less.
Divergence for $p^2$ is generated on boundaries
as in 
a four-dimensional case.

For $(p^2)^3$,
it is seen from Eqs.~(\ref{sp0}) and (\ref{firstapr}) 
that the integral to determine divergence 
in (\ref{p4b}) reduces to
$\int_0^\infty db \, e^{-Ab}$
where $A$ is a positive number.
This integration converges.
Similarly for $(p^2)^s (s\geq 3)$, 
the integrals converge.
This convergence is expected from an interpretation of 
a five-dimensional
integral as in Eq.~(\ref{k3d}).

Because of divergence of $(p^2)^2$ term,
the Lagrangian must have the counterterm. 
In next section, we will take into account this effect on
two-point functions.

\section{Bulk and boundary counterterms
and loop corrections \label{sec:counter}}

In this section, we examine quantum corrections
by starting with the action integral
including $(p^2)^2$ terms.

The action integral with $(p^2)^2$ is
\bea
  S_t=S+ S_{p^4} + S_{m^2} + S_{p^2} ,
\eea
In addition to the action $S$ given in Eq.~(\ref{action})
\bea
   S= \int d^4x \int_0^L dy \, 
    \left( {1\over 2} (\partial_\mu \phi)^2
     -{1\over 2}(\partial_y \phi)^2
     -{1\over 2}m^2 \phi^2
     -\lambda_5 \phi^4\right) ,
\eea
we take into account 
the corresponding terms 
to divergences appearing at two-loop level: 
\bea
  S_{p^4} 
  &\!\!\!=\!\!\!& \int d^4 x \int_0^L dy\,
    U\left(
     \partial_M \partial_N \phi \cdot
     \partial^M \partial^N \phi \right) ,
     \label{sp4}
\\
  S_{m^2} 
    &\!\!\!=\!\!\!& \int d^4x \int_0^L dy \,
      E\left[
      -{1\over 2} m^2 \phi^2
        +{1\over 8}
         \left(\partial_y^2 \phi\cdot \phi
           +\phi \partial_y^2 \phi
           -2 \partial_y \phi \partial_y \phi\right)
           \right]
        \delta_b,
    \label{sm2}
\\
  S_{p^2}&\!\!\!=\!\!\!&
    \int d^4x \int_0^L dy
     \, K \left[
       -{1\over 2}(\partial_\mu \phi)^2 \right]
       \delta_b ,
\eea
where $U$, $E$ and $K$ are parameters
and the sum of delta functions at the boundaries
is denoted as  
$\delta_b = \delta(y) +\delta(y-L)$.
The parameter $U$ corresponds to 
$-\lambda_5^2  /(35(2\pi)^4) \sim 
-2\times 10^{-5} \lambda^2 L^2$
in Eq.~(\ref{result1}).
The variation of these action integrals are given by
\bea
  \delta S &\!\!\!=\!\!\!&
  \int d^4 x \int_0^L dy
   \, \left[
   \left( -\partial_\mu \partial^\mu \phi
   +\partial_y^2 \phi
   -m^2 \phi
   -4\lambda_5 \phi^3\right) \delta \phi
  -\partial_y (\partial_y \phi\cdot \delta\phi)
   \right] ,
     \label{vl}
\\
   \delta S_{p^4}
 &\!\!\!=\!\!\!&
     \int d^4x \int_0^L dy \,
  2U\left[
   (\partial_\mu \partial^\mu - \partial_y)^2 
 \phi \cdot \delta \phi
   \right.
\nonumber
\\
   && \left.
   +2 \partial_y \left((\partial_\mu\partial^\mu
    -\partial_y^2)\partial_y \phi\cdot
   \delta\phi\right)
  +\partial_y^2
   (\partial_y^2 \phi\cdot \delta\phi) 
  \right] , 
\\
   \delta S_{m^2}
  &\!\!\!=\!\!\!&
   \int d^4x \int_0^L dy
 \,
  E\left[
   (-m^2 +\partial_y^2) \phi \cdot
    \delta\phi  \delta_b    
  \right.
\nonumber
\\
  && \!\!\!\!\!
    +{1\over 4}
      \partial_y \left(
      \phi \partial_y \delta \phi \cdot
      \delta_b
      -3\partial_y \phi\cdot \delta \phi
       \delta_b
      -\phi \cdot \delta\phi \partial_y \delta_b\right)
  \left.
    +{1\over  4}
    \left(
      4\partial_y \phi\cdot \partial_y \delta_b
      + \phi \partial_y^2 \delta_b \right)
    \cdot \delta\phi
    \right] ,
    \label{ve}
\\
   \delta S_{p^2}
     &\!\!\!=\!\!\!&
     \int d^4x \int_0^L dy
     \,
     K\left[ \partial_\mu \partial^\mu \phi\cdot
      \delta \phi\right]
     \delta_b .
     \label{vk}
\eea
The effect of the boundary kinetic terms (\ref{vk}) 
on the leading terms (\ref{vl}) makes
mixing of 
Neumann and Dirichlet conditions as boundary
conditions~%
\cite{Csaki:2003dt}.
The contributions
for Eqs.~(\ref{vk}) and (\ref{vl})
can also be treated in a way based on hermitian
property of differential operators~%
\cite{Csaki:2005vy}.
The effect of Eq.~(\ref{ve}) is similarly 
restrictive such that
the existence of zero mode is not guaranteed~%
\cite{delAguila:2003bh}.
Thus, these boundary terms change the solutions for
mode functions.
Instead of solving full solutions,
we can reasonably assume the existence of the solutions
as the boundary terms are regarded as perturbation.

From the above equations,
bulk quadratic equation of motion is
\bea
   (-\partial_\mu \partial^\mu
    +\partial_y^2 -m^2
    + 2U (\partial_\mu \partial^\mu 
    -\partial_y^2)^2)\phi= 0 .
\eea
By Fourier transformation, this equation is
written as
\bea
   (p^2 -m_n^2 -m^2
     +2U (p^2 -m_n^2)^2 )\phi_n = 0 .
\eea
The propagator is obtained as
\bea
   {i\over p^2 -m_n^2 -m^2
     +2U (p^2 -m_n^2)^2}
  = {i\over 2U\left[ \left(
     p^2 +{1\over 4U}
        -m_n^2 \right)^2
     -{1\over 16U^2}
        (1+8Um^2) \right] }.
        \label{prop4}
\eea
For $1+8Um^2 >0$, Eq.~(\ref{prop4}) is
\bea
    && 
  {1\over \sqrt{1+8Um^2}}
   \left(
   {i\over  p^2 +{1-\sqrt{1+8Um^2} \over 4U}
        -m_n^2 }
    -
   {i\over  p^2 +{1+\sqrt{1+8Um^2} \over 4U}
        -m_n^2} \right)  .
     \label{propp}
\eea
For the poles of the propagators,
positive masses squared
\bea
    {1- \sqrt{1+8Um^2}\over 4U} < 0 ,\qquad
    {1+\sqrt{1+8Um^2}\over 4U} < 0,
\eea 
correspond to $U<0$.
In Eq.~(\ref{propp}), the second term
has the unusual sign so that their degrees
of freedom should decay.
This requires hierarchy between masses
\bea
   {1-\sqrt{1-8|U|m^2}\over 4|U|}
  \ll 
    {1+\sqrt{1- 8|U|m^2}\over 4|U|} ,
\eea
where the mass of the lighter degree of
freedom is at least of order $m_n$ and $U<0$.
Then the propagator (\ref{propp}) 
is approximately given by
\bea
   {i\over  p^2 -m^2 -m_n^2 }
    -
   {i\over  p^2 -{1\over 2|U|} -m_n^2}  .
    \label{twopoles}
\eea
The propagator has two poles
for the masses squared $m^2 +m_n^2$
and ${1\over 2|U|} +m_n^2$.
%
The new pole in Eq.~(\ref{twopoles}) is directly related to
the $(p^2)^2$ action (\ref{sp4}).
The corresponding counterterms can be employed
to bring in a value at a point of energy scales.
For example, it would be possible to set 
the new pole term to zero at one scale.
In this case, the value can become nonzero at other scales.
In order to make the undesirable mode decay, 
this value must be small.
If the concept of an effective theory is fulfilled 
as in a four-dimensional theory where
higher derivative operators are irrelevant,
the coefficient may be kept small at low energies.
To make a general statement on the running 
of higher-dimensional operators is beyond the scope
of the paper.
When the coefficient for the $(p^2)^2$ term
needs to be specified in the following,
we will adopt the size we have found at two-loop level.

Now we evaluate the effect of the two poles
for two-point functions.
From Eq.~(\ref{1cutoffint}),
the boundary integral
in the one-loop two-point function (\ref{eval}) is
\bea
 &&  \int {d^4 p \over (2\pi)^4} 
    \left({i\over p^2 -m^2 -m_n^2}
  -{i\over p^2 +{1\over 2U} -m_n^2}\right)
\nonumber
\\
  &\!\!\!=\!\!\!&
  {1\over 16\pi^2}
 \left( m_n^2 \log {m_n^2 +m^2 \over
   m_n^2 -{1\over 2U}}
 -{1\over 2U} \log
    {\Lambda^2\over m_n^2 -{1\over 2U}}
  -m^2 \log {\Lambda^2 \over
    m_n^2 +m^2}\right)
    . \label{beforea}
\eea  
The effect of two poles makes the divergence
on boundaries logarithmic.
The dominant contribution for the divergence
comes from ${1\over 2|U|} \log \Lambda^2$.
When counterterms are taken into account,
corrections are described as
${1\over 2|U|} \log \mu^2$,
where $\mu$ is a running scale.
For $U$ as a loop effect
such as $|U|\sim 2\times 10^{-5} \lambda^2 L^2$,
the size of $1/U$ depends on $\Lambda L$,
\bea
   {1\over 2|U|} \sim {10^2 \Lambda^2 \over \lambda^2}
    > \Lambda^2 , \quad \textrm{for} \quad
      \Lambda L=10 ,
\qquad
  {1\over 2|U|}
  \siml \Lambda^2 ,
   \quad \textrm{for}\quad  \Lambda L\simg 10^2 .
   \label{eval1}
\eea
For $\Lambda L=10$,
the degree of freedom with a mass
larger than the cutoff scale is needed.
This is a breakdown of the model.
For $\Lambda L \simg 10^2$,
the coefficient can be parameterized
as ${1\over 2|U|}=a\Lambda^2$, where
$a\siml 1$.
Then Eq.~(\ref{beforea}) reduces to
\bea
   {1\over 16\pi^2}
   \left( -m_n^2 
    -(m_n^2 +m^2) \log {\Lambda^2 \over m_n^2 +m^2}
    +a\Lambda^2 \log {1\over a} 
    \right) .
\eea
This equation means that
the physical quantity is extracted
with a fine tuning
as in a four-dimensional $\phi^4$ theory
and that
the fine tuning can be moderate depending
on $a$.

The bulk divergent terms
in the one-loop two-point function (\ref{eval})
are
\bea
&& \left.
    \sum_{n=-\infty}^\infty
     \int {d^4k\over (2\pi)^4}
     \left({i\over k^2 -m^2 -m_n^2}
     -{i\over k^2 +{1\over 2U} -m_n^2}
   \right) \right|_{\textrm{\scriptsize div}}
\nonumber
\\
  &\!\!\! = \!\!\!&
  {1\over 2^4 \pi^{5\over 2}}
    \left[
  \left({2\over 3}\Lambda^2
     -2m^2 \right) \Lambda L
  -\left({2\over 3}\Lambda^2 
   +{1\over U}\right) \Lambda L \right]
\nonumber
\\
  &\!\!\!=\!\!\!& 
  {1\over 2^4 \pi^{5\over 2}}
  \left(
     -2m^2
  - {1\over U}\right)  \Lambda L .
\eea  
Replacing a cutoff regularization by
a dimensional regularization and taking into account counterterms leads to
the correction
\bea
  {1\over 2^4 \pi^{5\over 2}}
  \left(
     -2m^2
  - {1\over U}\right)  \log \mu  .
\eea
The dominant coefficient is
\bea
   {1\over 2^4 \pi^{5\over 2}} {1\over |U|}
  \sim    
  {1\over 2^4 \pi^{5\over 2}} 
    {10^5\over 2\lambda^2 L^2} 
  \sim {100 \over \lambda^2 (\Lambda L)^2} \Lambda^2
  \siml \Lambda^2 \qquad \textrm{for}
  \quad \Lambda L \simg 10^2 .
\eea
The physical quantity can be extracted
for a large cutoff
$\Lambda > 10^2 L^{-1}$ similarly to the case of
boundary divergent terms.

\section{Higher derivative corrections beyond two loops
\label{sec:high}}

In this section, we discuss higher derivative corrections
beyond two-loop level.
By examining
the structure of operators for two and three poles,
we make a conjecture on a general form for two-point functions
with multiple poles. 
In addition, we give an example of a counterterm for $(p^2)^3$.

We consider the propagator with three poles which 
has the form
\bea
    {A\over p^2 -m_1^2} 
     +{B\over p^2 -m_2^2} 
      +{C\over p^2 -m_3^2}
    ={F\over (p^2 -m_1^2)(p^2-m_2^2)(p^2-m_3^2)} ,
    \label{histart}
\eea
where the poles are located at $p^2=m_i^2$, $i=1,2,3$.
The constants $A$, $B$, $C$, $F$ are independent of $p^2$
as the right-hand side is the inverse of
the sum of operators with $(p^2)^n$, $n=1,2,3$.
The four constants satisfy
\bea
  && A+B+C=0 ,
  \label{heq1}
\\
  && 
   A(m_2^2 +m_3^2)
    +B(m_3^2 +m_1^2)
     +C(m_1^2 +m_2^2) =0 ,
   \label{heq2}
\\
  && Am_2^2 m_3^2
   +Bm_3^2 m_1^2 
     +Cm_1^2 m_2^2 = F .
  \label{heq3}
\eea
The equation (\ref{heq1}) means $C=-A-B$.
Substituting this equation into Eq.~(\ref{heq2}) yields
\bea
   A(m_3^2 -m_1^2) +B(m_3^2 -m_2^2) = 0 .
   \label{heq2v}
\eea
The solutions for this equation are classified for $m_3^2 \neq m_2^2$
and for $m_3^2 =m_2^2$.
For the case $m_3^2 = m_2^2$,
the left-hand side in Eq.~(\ref{histart}) is
$A(m_1^2-m_3^2)/[(p^2-m_1^2)(p^2-m_3^2)]$
and there is no solution.
For the case $m_3^2 \neq m_2^2$,
Eq.~(\ref{heq2v}) has the solution
$B=-[(m_3^2-m_1^2)/(m_3^2 -m_2^2)] A$.
Substituting this equation and $C=-A-B$
into Eq.~(\ref{heq3}) yields $A(m_2^2- m_1^2)(m_3^2-m_1^2) =F$.
Because $F\neq 0$, $A$ is 
$A= F/[(m_2^2 -m_1^2)(m_3^2 -m_1^2)]$.
In this case, $m_1^2$, $m_2^2$ and $m_3^2$
need to be different from each other.
Without loss of generality, these masses can be taken 
as $m_1^2<m_2^2 <m_3^2$.
The inverse of the equation (\ref{histart}) becomes
\bea
   \left[{F\over (p^2-m_1^2)
   (p^2-m_2^2)(p^2-m_3^2)} \right]^{-1}
  &\!\!\!=\!\!\!&{1\over F}(p^2)^3
    -{m_1^2+m_2^2+m_3^2\over F}(p^2)^2
\nonumber
\\
 &&
  +{m_1^2 m_2^2 +m_2^2 m_3^2 +m_3^2 m_1^2 \over F}p^2
  -{m_1^2 m_2^2 m_3^2 \over F} .
   \label{inverse1}
\eea
Because the quadratic term is an usual kinetic energy term,
the coefficient for $p^2$ can be fixed as
$(m_1^2 m_2^2  +m_2^2 m_3^2 + m_3^2 m_1^2)/F
=1$.
Then $F>0$.
The coefficient of the propagator for the lightest particle is 
$A=F/[(m_2^2 -m_1^2)(m_3^2-m_1^2)] > 0$.
The particle with the pole 
at $p^2=m_2^2$ has a negative coefficient 
$B=F/[(m_1^2 -m_2^2)(m_3^2 -m_2^2)] <0$.
Therefore this particle must decay. 
The mass needs to be large, $m_2^2 \gg m_1^2$.
Because $m_3^2 >m_2^2$,
the masses are hierarchical for the lightest mode and
the others,
$m_1^2 \ll m_2^2 , m_3^2$.
Then $F\sim m_2^2 m_3^2$.
The equation (\ref{inverse1}) is obtained as
\bea
  {1\over m_2^2 m_3^2} (p^2)^3
    -\left({1\over m_2^2} +{1\over m_3^2}\right)
      (p^2)^2 + p^2 -m_1^2 .
      \label{inverse2}
\eea
The propagator with the three poles is written as
\bea
   {1\over p^2 -m_1^2}
   -\left({m_3^2\over m_3^2 -m_2^2}\right) {1\over p^2 -m_2^2}
   +\left({m_2^2 \over m_3^2 -m_2^2}\right) {1\over p^2 -m_3^2} .
  \label{propni2}
\eea
From the solution~(\ref{inverse2}) for $(p^2)^3$ and 
the $(p^2)^2$ term,
we expect a generalized form
\bea
  && p^2 -m_1^2 
   -\left({1\over m_n^2} +{1\over m_{n-1}^2}
  +\cdots + {1\over m_2^2}\right) (p^2)^2
\nonumber
\\
  &&  +\left(
  {1\over m_n^2 m_{n-1}^2}
  +{1\over m_{n-1}^2 m_{n-2}^2}
   +\cdots + {1\over m_3^2 m_2^2} \right) (p^2)^3
\nonumber
\\
  && -\left(
  {1\over m_n^2 m_{n-1}^2 m_{n-2}^2}
  +\cdots + {1\over m_4^2 m_3^2 m_2^2} \right) (p^2)^4 
 + \cdots
  +(-1)^{n+1}
  {1\over m_n^2 \cdots m_2^2} (p^2)^n ,
\eea
for $m_1^2 \ll m_2^2 < \cdots < m_n^2$.
The low-energy predictability can be treated 
as in the case of $(p^2)^2$
for $m_1\ll m_2$.

\subsubsection*{Four-loop diagram}

\begin{figure}[htb]
\begin{center}
\includegraphics[width=6cm]{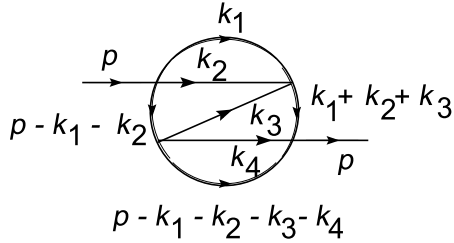}
\caption{A four-loop diagram for $(p^2)^3$-divergence.
 \label{fig:4loop}}
\end{center}
\end{figure}

We here discuss how divergence of
higher derivative terms than $(p^2)^2$ appears. 
From the dimensional analysis (\ref{diman}),
divergence of $(p^2)^3$ terms 
is expected to arise from four-loop diagrams.
We consider a four-loop diagram shown in Figure~\ref{fig:4loop}.
In Figure~\ref{fig:4loop},
four-momenta are drawn and
the contributions of Kaluza-Klein mode should be summed.
In order to estimate the appearance of bulk divergence,
we employ an approximation for the loop-integral 
and the mode summation, following the interpretation
in Eqs.~(\ref{45to}) and (\ref{k3d}), as
\bea
    \sum_n \int {d^4 k\over (2\pi)^4}
    \to \int {d^5 \bar{k} \over (2\pi)^5} .
\eea 
Here $\bar{k}$ is a five-dimensional momentum which 
expresses a four-dimensional momentum and 
a Kaluza-Klein mass collectively.
The bulk divergence diagram shown in Figure~\ref{fig:4loop}
is approximately obtained from a five-dimensional calculation
\bea
  &&   (-i\lambda_5)^4
     \int {d^5 \bar{k}_1\over (2\pi)^5}
   {d^5 \bar{k}_2\over (2\pi)^5}
{d^5 \bar{k}_3\over (2\pi)^5}
{d^5 \bar{k}_4\over (2\pi)^5}
\nonumber
\\  
  && \times
  {i\over \bar{k}_1^2}   
   {i\over \bar{k}_2^2}   
   {i\over \bar{k}_3^2}   
  {i\over \bar{k}_4^2}  
  {i\over (\bar{k}_1 +\bar{k}_2 +\bar{k}_3)^2}
  {i\over (\bar{p}-\bar{k}_1
     -\bar{k}_2)^2}  
  {i\over (\bar{p}- \bar{k}_1
      -\bar{k}_2 -\bar{k}_3
      -\bar{k}_4)^2}  ,
\eea
which is multiplied by a symmetric factor $(4!)^5$.
The number of the momentum integral is 4
and the number of the propagator is 7.
The structure of divergence
is described in power series with respect to $(p^2)$ as 
\bea
   -i\lambda_5^4 \sum_{n=0}^\infty
    C_n \Lambda^{4\times 5 - 7\times 2 -2n}(p^2)^n
  &\!\!\!=\!\!\!&
   -i\lambda^4 \left[(\Lambda L)^4 
 \left(C_0  \Lambda^2
  + C_1 p^2\right) 
  + C_2 (\Lambda L)^2 (p^2)^2 L^2 
  \right.
\nonumber
\\
  && \left.
  + C_3 \log (\Lambda L) (p^2)^3 L^4 \right]
  + (\textrm{finite})  ,
  \label{contri}
\eea
which shows that
the $(p^2)^3$ term has logarithmic divergence.
The coefficients $C_3$ is estimated as
order of
${\cal O}((4!)^5 (8\pi^2 /(3(2\pi)^5))^4) \sim {\cal O}(10^{-4})$ 
where the area of a five-dimensional unit sphere is
$\int d\Omega_5 ={8\pi^2 /3}$.
With the dimensionless quantity $(\Lambda L)$
for the scale $L$,
Eq.~(\ref{contri}) is rewritten as
\bea
   -i\lambda^4 (\Lambda L)^4  \left[
    C_0  \Lambda^2
  + C_1 p^2 
  + C_2 { (p^2)^2 \over \Lambda^2} 
  + C_3 \log (\Lambda L) {(p^2)^3 \over \Lambda^4} \right]
  + (\textrm{finite})  .
\eea
The corresponding 
two-loop contribution is obtained from Eq.~(\ref{result1}).
Combining the two and three-loop contributions,
the two large masses in the three-pole expression (\ref{inverse2})
are estimated as
\bea
   m_2^2 \sim 10^4 {\Lambda^2 \over (\lambda\Lambda L)^4} ,
 \qquad
   m_3^2 \sim {\Lambda^2 \over \log(\Lambda L)} ,
\eea
for $10^{-1}\log(\Lambda L)/(\lambda \Lambda L)^2<1$
whose opposite inequality seems an unnatural option.
Similarly to the evaluation in Eq.~(\ref{eval1}),
this also shows that the ultraviolet cutoff 
must be orders of magnitude larger compared to
a compactification scale.

\section{Conclusion \label{sec:concl}}

We have studied quantum loop corrections to
two-point functions and extraction of physical
quantities in a five-dimensional 
$\phi^4$ theory on an orbifold $S^1/Z_2$.
As in four dimensions, divergence
for mass terms is found
at one-loop and
divergence for wave functions 
is found at two-loop level.
At the two-loop level,
we have found 
the divergence for quartic derivative terms 
of $(p^2)^2$ that requires 
counterterms in the bulk.
These terms make
propagators have two poles.
One of the degrees of freedom 
for the two poles
has the 
unusual sign for the propagator and
should be heavy so that it decays.
With this effect taken into account,
corrections to two-point functions have been derived. 
The contributions of the quartic derivative
terms to masses depend on both of an ultraviolet
cutoff $\Lambda$ and
a size of the extra dimension $L$.
We have shown that the contributions can be
extracted with a moderate fine-tuning for 
$\Lambda L \simg 10^2$ and 
that they cannot be fixed for
$\Lambda L \sim 10$.
Therefore,
for extraction of physical quantities
for two-point functions an
ultraviolet cutoff must be orders of
magnitude larger compared to a compactification
scale.
%
This behavior is found for even higher correcions with
multiple poles at higher loop levels.

Since divergence of higher derivative terms 
appears in various 
models with compactified extra dimensions,
counterterms must be included.
If a gauge field propagates in the bulk,
the gauge field $A_\mu$ has mass dimensions
$[A_\mu]= [\textrm{mass}]^{3\over 2}$ and
the gauge coupling
$g_A$ has mass dimensions
$[g_A] = [\textrm{mass}]^{-{1\over 2}}$.
Possible Lagrangian terms for 
counterterms are
\bea
   (\partial A)^2 ,\quad
   g_A (\partial A) A^2 ,\quad
   g_A^2 A^4 ,\quad
   g_A^4 (\partial^2 A)^2 ,\quad    
   g_A^5 (\partial A)^3 ,\quad 
   g_A^6 (\partial A)^2 A^2 ,\quad \cdots,
\eea
where vector indices are contracted.
In this case, from the requirement of 
extracting physical quantities
with higher derivative terms
the value of $\Lambda L$ may be 
constrained.
In flat five dimensions, 
dependence of the gauge coupling on energy is
usually large compared to the case of four dimensions~%
\cite{ddg,run}.
For a large $\Lambda L$, models could give 
rise to  
$g_A =0$ or $g_A \gg 1$.
Therefore the effects of energy dependence of 
gauge couplings and higher derivative terms
should be carefully treated so that
models are consistently formulated.

\vspace{8ex}

\subsubsection*{Acknowledgments}

I thank Yutaka Hosotani for helpful comments
and suggestions
about treatment of quantum corrections
in models with extra dimensions.
This work is supported by Scientific Grants 
from the Ministry of Education
and Science, Grant No.~20244028.

\newpage

\begin{appendix}
 
\section{Calculations of one-loop diagrams
\label{ap:c1}}

From Eq.~(\ref{actionk}), 
the interaction Lagrangian is
\bea
   {\cal L}_{int}
     ={\cal L}_1 + {\cal L}_2 
       +{\cal L}_3 + {\cal L}_4 ,
       \label{lintap}
\eea
with expansion in terms of $\phi_0$ and $\phi_{|n|}$ as
\bea
   {\cal L}_1 &\!\!\!=\!\!\!&
       -\lambda_1 \phi_0^4 ,
\qquad
  {\cal L}_2 =
      -\lambda_2 \phi_0^2 \phi_{|n|}^2 ,
\qquad
  {\cal L}_3 =
     -\lambda_3 \phi_0 \phi_{|n|}
      \phi_{|\ell|}
       \phi_{|n+\ell|} ,
\\
  {\cal L}_4 &\!\!\!=\!\!\!&
    -\lambda_4 \phi_{|n|}
     \phi_{|\ell|}
      \phi_{|s|}\phi_{|n+\ell+s|} .
\eea
Here the coupling constants are
\bea
   \lambda_1 &\!\!\!=\!\!\!&
      {\lambda_5 \over 4L} \, \left(
         (\sqrt{2}-1)^4+ 4(\sqrt{2} -1)^3\right) ,
\\
  \lambda_2 &\!\!\!=\!\!\!&
    {\lambda_5 \over 4L} \, 6(\sqrt{2}-1)^2 ,
\qquad
  \lambda_3 =
    {\lambda_5 \over 4L}
      \, 4(\sqrt{2}-1) ,
\qquad
  \lambda_4 =
    {\lambda_5 \over 4L} .
\eea
For the two-point function 
with the external lines $\phi_0\phi_0$,
The contributions from each part of
Lagrangian (\ref{lintap}) are as follows:
\bea
  \begin{picture}(80,30)
  \put(0,-10){\includegraphics[width=2cm]{1-loop.eps}}
  \put(-10,-10){0}
  \put(60,-10){0}
  \put(23,4){${\cal L}_1$}
  \end{picture}
  &\!\!\!=\!\!\!&
  -12 i\lambda_1 {\cal D} ,
\\
  \begin{picture}(80,35)
  \put(0,-10){\includegraphics[width=2cm]{1-loop.eps}}
  \put(-10,-10){0}
  \put(60,-10){0}
  \put(23,4){${\cal L}_2$}
  \end{picture}
 &\!\!\!=\!\!\!& 
  -2 i\lambda_2 \sum_{n=-\infty}^\infty
    {\cal D}_n
  -10i\lambda_2  {\cal D},
\\
  \begin{picture}(80,30)
  \put(0,-10){\includegraphics[width=2cm]{1-loop.eps}}
  \put(-10,-10){0}
  \put(60,-10){0}
  \put(23,4){${\cal L}_3$}
  \end{picture}
  &\!\!\!=\!\!\!& -6 i\lambda_3 \sum_{n=-\infty}^\infty
    {\cal D}_n
  -6i\lambda_3  {\cal D} ,
\\
  \begin{picture}(80,30)
  \put(0,-10){\includegraphics[width=2cm]{1-loop.eps}}
  \put(-10,-10){0}
  \put(60,-10){0}
  \put(23,4){${\cal L}_4$}
  \end{picture}
  &\!\!\!=\!\!\!& -12 i\lambda_4 \sum_{n=-\infty}^\infty
    {\cal D}_n .
\eea
The total contribution is obtained as
\bea
  \begin{picture}(80,30)
  \put(0,-10){\includegraphics[width=2cm]{1-loop.eps}}
  \put(-10,-10){0}
  \put(60,-10){0}
  \put(23,4){${\cal L}_{int}$}
  \end{picture}
  &\!\!\!=\!\!\!&  
 -6i\lambda {\cal D}
    -6i\lambda \sum_{n=-\infty}^\infty
     {\cal D}_n  .
    \label{1-00}
\eea
Here
\bea
 2(6\lambda_1 +5\lambda_2 +3\lambda_3)
  ={6\lambda_5\over L} ,
\qquad
 2(\lambda_2 +3\lambda_3 +6\lambda_4)
    ={6\lambda_5 \over L} .
\eea

For two-point functions with the external lines $\phi_{|f|}\phi_{|g|}$, one-loop contributions 
from each diagram are  
\bea
  \begin{picture}(80,30)
  \put(0,-10){\includegraphics[width=2cm]{1-loop.eps}}
  \put(-15,-10){$|f|$}
  \put(60,-10){$|g|$}
  \put(23,4){${\cal L}_1$}
  \end{picture}
  &\!\!\!=\!\!\!&
  -12 i\lambda_1 \delta_{f0} \delta_{g0}
  {\cal D} ,
\\
  \begin{picture}(80,30)
  \put(0,-10){\includegraphics[width=2cm]{1-loop.eps}}
  \put(-15,-10){$|f|$}
  \put(60,-10){$|g|$}
  \put(23,4){${\cal L}_2$}
  \end{picture}
  &\!\!\!=\!\!\!& -2 i\lambda_2 \, \delta_{f0} \delta_{g0}
  \sum_{n=-\infty}^\infty
    {\cal D}_n
\nonumber
\\
  &&
  -8i\lambda_2 \,\delta_{f0} \delta_{g0}
    {\cal D}
  -2i\lambda_2 \,\delta_{|f||g|}
 {\cal D} ,
\\
  \begin{picture}(80,30)
  \put(0,-10){\includegraphics[width=2cm]{1-loop.eps}}
  \put(-15,-10){$|f|$}
  \put(60,-10){$|g|$}
  \put(23,4){${\cal L}_3$}
  \end{picture}
  &\!\!\!=\!\!\!& -6 i\lambda_3 \, \delta_{f0} \delta_{g0}
  \sum_{n=-\infty}^\infty
    {\cal D}_n
\nonumber
\\
  && -3 i\lambda_3 \, \delta_{f0} 
  \sum_{n=-\infty}^\infty \delta_{|g||2n|}
    {\cal D}_n
   -3 i\lambda_3 \,  \delta_{g0}
  \sum_{n=-\infty}^\infty \delta_{|f||2n|}
    {\cal D}_n
\nonumber
\\
  &&
  +6i\lambda_3 \,\delta_{f0} \delta_{g0}
    {\cal D}
  -6i\lambda_3 \,\delta_{|f||g|}
  {\cal D} ,
\\
  \begin{picture}(80,30)
  \put(0,-10){\includegraphics[width=2cm]{1-loop.eps}}
  \put(-15,-10){$|f|$}
  \put(60,-10){$|g|$}
  \put(23,4){${\cal L}_4$}
  \end{picture}
  &\!\!\!=\!\!\!& -12 i\lambda_4 \, \delta_{|f||g|}
  \sum_{n=-\infty}^\infty
    {\cal D}_n
\nonumber
\\
  && -12 i\lambda_4 
  \sum_{n=-\infty}^\infty \delta_{|f||g+2n|}
    {\cal D}_n
  +12i\lambda_3 \,\delta_{|f||g|}
  {\cal D} .
\eea
The sum of the contributions is
\bea
   \begin{picture}(80,30)
  \put(0,0){\includegraphics[width=2cm]{1-loop.eps}}
  \put(-15,0){$|f|$}
  \put(60,0){$|g|$}
  \put(23,14){${\cal L}_{int}$}
  \end{picture}
  &\!\!\!=\!\!\!&
 - i \delta_{f0} \delta_{g0}
  {\cal D} 
     (12\lambda_1 +8\lambda_2 -6\lambda_3)
\\
  && - i \delta_{f0} \delta_{g0}
  \sum_{n=-\infty}^\infty
    {\cal D}_n
   (2\lambda_2 + 6\lambda_3)
\nonumber
\\
  &&
  -i\delta_{|f||g|}
  {\cal D}
  (2\lambda_2 +6\lambda_3 -12\lambda_4) 
\nonumber
\\
  && -3 i\lambda_3 \, \delta_{f0} 
  \sum_{n=-\infty}^\infty \delta_{|g||2n|}
    {\cal D}_n
   -3 i\lambda_3 \,  \delta_{g0}
  \sum_{n=-\infty}^\infty \delta_{|f||2n|}
    {\cal D}_n
\nonumber
\\
  && -12 i\lambda_4 \, \delta_{|f||g|}
  \sum_{n=-\infty}^\infty
    {\cal D}_n
   -12 i\lambda_4 
  \sum_{n=-\infty}^\infty \delta_{|f||g+2n|}
    {\cal D}_n
 .
  \label{tfg1}
\eea
With the values,
\bea
  && 2(6\lambda_1 +4\lambda_2 -3\lambda_3)
  = 3(\sqrt{2}-1)^2 \lambda ,
\qquad
   2(\lambda_2 +3\lambda_3) 
   = 3\lambda ,
\\
  && \lambda_2 +3\lambda_3 -6\lambda_4
     = 0 ,
\eea
the amplitude~(\ref{tfg1}) becomes Eq.~(\ref{eq1loopt}).
For $f=g=0$, Eq.~(\ref{eq1loopt})
reduces to Eq.~(\ref{1-00}).

\section{Divergent integrals \label{ap:dint}}

In this section, formulas for 
various divergent integrals are presented.

An integral often appearing in orbifold models is
\bea
   \sum_{n=-\infty}^\infty \int
     {d^4k\over (2\pi)^4} \,
     {i\over \left[k^2 -m^2 -m_n^2\right]^s} 
  &\!\!\!=\!\!\!&
  \sum_{n=-\infty}^\infty \int
     {d^d k_E\over (2\pi)^d} \,
     {(-1)^{s+1} \over 
 \left[k_E^2 +m^2 +m_n^2\right]^s}     
\nonumber
\\
  &\!\!\!=\!\!\!&    
      \sum_{n=-\infty}^\infty
        \int {d^d k_E\over (2\pi)^d} \,
        {(-1)^{s+1}\over \Gamma(s)}
        \int_0^\infty dt \, t^{s-1}
        e^{-(k_E^2 + m^2 + {n^2\pi^2\over L^2})t} 
\nonumber
\\
   &\!\!\!=\!\!\!&
   \sum_{n_p=-\infty}^\infty
     {L\over 2^d \pi^{d+1\over 2}}
     {(-1)^{s+1}\over \Gamma(s)}
     \int_0^\infty dt \, t^{s-{d\over 2}-{3\over 2}}
     e^{-m^2 t -{n_p^2 L^2 \over t}} .
     \label{integ1}
\eea
In the first equality, 
$4\to d=4-\epsilon$ and $k^0 = ik_E^0$ have 
been employed. 
In the second equality, we have applied
an identity for Gamma function 
and Poisson's summation formula
\bea
     {1\over A^s}
       ={1\over \Gamma(s)}
        \int_0^\infty dt \, t^{s-1} e^{-At} ,        
\qquad 
    \sum_{n=-\infty}^\infty
      e^{-b(n-a)^2}
      =\sqrt{\pi\over b}
      \sum_{n_p=-\infty}^\infty
      e^{-{\pi^2 n_p\over b}
      -2\pi i n_p a} .
\eea
In the last equality in Eq.~(\ref{integ1}),
Gaussian integral has been performed.
With a formula for a modified Bessel function,
\bea
   \int_0^\infty dt \, t^{\eta-1}
     e^{-a^2 t} e^{-{b^2\over t}}
      =2\left({b\over a}\right)^\eta
        K_{-\eta} (2ab) ,
\eea
for $\eta<{1\over 2}$,
Eq.~(\ref{integ1}) is
\bea
 && 
     {L\over 2^d \pi^{d+1\over 2}}
     {(-1)^{s+1}\over \Gamma(s)}
     \left[
     \int_0^\infty dt \, t^{s-{d\over 2}-{3\over 2}}
     e^{-m^2 t}
  +
     \sum_{n_p=1}^\infty
   2^2 \left({n_p L\over m}\right)^{s-{d+1\over 2}}
   K_{{d+1\over 2}-s}
   (2n_p Lm) \right]  .
    \label{sgen}
\eea

For $s=1$, Eq.~(\ref{sgen}) is
\bea
  \sum_{n=-\infty}^\infty {\cal D}_n
  &\!\!\!=\!\!\!& 
  {L\over 2^4 \pi^{5\over 2}}
     \int_0^\infty dt \, t^{-{5\over 2}}
     e^{-m^2 t} 
  + \sum_{n_p=1}^\infty
     {L\over 2^2 \pi^{5\over 2}}
    \left({m\over n_p L}\right)^{{3\over 2}}
   K_{{3\over 2}}
   (2n_p Lm)   
\nonumber
\\
  &\!\!\!=\!\!\!&
  {L\over 2^4 \pi^{5\over 2}}
     \int_0^\infty dt \, t^{-{5\over 2}}
     e^{-m^2 t} 
  +  \sum_{n_p =1}^\infty
      {1\over (2\pi)^2}
      {m\over 2L}
      \, {1\over n_p^2}
      \left(1+{1\over 2n_p Lm}\right)
      e^{-2n_p Lm} ,
       \label{ints1}
\eea  
where
\bea
  K_{3\over 2}(z) =\sqrt{\pi \over 2z}
    \left(1+{1\over z}\right) e^{-z} .
\eea
In Eq.~(\ref{ints1}),
the first term diverges at $t\to 0$.
With a cutoff near $t\sim 0$, the divergent part of 
the integral is obtained as
\bea
  \int_0 dt \, 
    t^{-{5\over 2}}(1-m^2 t+ \cdots)
  &\!\!\!=\!\!\!&
  \left[
   -{2\over 3} t^{-{3\over 2}}
  -m^2 (-2) t^{-{1\over 2}} +\cdots 
  \right]_{t={1\over \Lambda^2}}    
\nonumber
\\
  &\!\!\! =\!\!\!&
   {2\over 3}\Lambda^3
      -2m^2\Lambda .
\eea
From these equations,
divergence in Eq.~(\ref{bulkterm}) becomes
Eq.~(\ref{1bulk}).

Equations for integrals at two-loop level
are given in the following. 

Eq.~(\ref{p4b}) includes the integral
\bea
  && \sum_{s=-\infty}^\infty
     \int {d^4 k_2 \over (2\pi)^4}
      \, {1\over k_2^2 -m_s^2}
      {1\over ((k_1 +k_2)^2 -m_{f+\ell+s}^2)^h}
\nonumber
\\
 &\!\!\!=\!\!\!&
  \sum_{s=-\infty}^\infty
    \int_0^1 dx
    \int{d^4k_2 \over (2\pi)^4}
    {hx^{h-1}
    \over 
    \left[(1-x)
    \left(k_2^2 - {s^2\pi^2 \over L^2}\right)
    +x
    \left((k_1+k_2)^2 
     -{(f+\ell+s)^2 \pi^2 \over L^2}\right)
     \right]^{h+1}}
\nonumber
\\
  &\!\!\!=\!\!\!&
   \sum_{s=-\infty}^\infty
    \int_0^1 dx \int {d^4 \ell \over 
    (2\pi)^4}
      \, {hx^{h-1}\over \left[
      \ell^2 -\Delta\right]^{h+1}} ,
    \label{dh}  
\eea
with a positive number $h$.
Here
\bea
  \ell =
  k_2 +x k_1 ,
 \qquad 
 \Delta  =
 x(1-x) \left(-k_1^2 +m_{f+\ell}^2
   \right)
  +{(s+x(f+\ell))^2 \pi^2 \over L^2} .
\eea
Employing Gamma function and Poisson's summation formula
similarly to derivation in Eq.~(\ref{integ1}),    
Eq.~(\ref{dh}) is
\bea
  && \sum_{s=-\infty}^\infty
    \int_0^1 dx \int {d^4 \ell_E \over 
    (2\pi)^4}
      \, hx^{h-1} {1\over \Gamma(h+1)}
    \int_0^\infty dt \, t^h 
 e^{-(\ell_E^2 +\Delta)t}
 i(-1)^{h+1}
\nonumber
\\
  &\!\!\!=\!\!\!&
  \sum_{s_p=-\infty}^\infty
  \int_0^1 dx 
   {i(-1)^{h+1} L\over 2^d \pi^{d+1\over 2}}
  {hx^{h-1}\over \Gamma(h+1)}
   \int_0^\infty dt \, t^{h-{d+1\over 2}}
\nonumber
\\
  && \times
   \exp\left[-x(1-x)\left(-k_1^2 
 +m_{f+\ell}^2 \right)t
  -{s_p^2 L^2 \over t}
  +2\pi i s_p x(f+\ell) \right] .
   \label{sp2}
\eea
In Eq.~(\ref{sp2}), divergence arises from terms 
for $s_p=0$.
For $s_p=0$, Eq.~(\ref{sp2}) is
\bea
  \int_0^1 dx 
   {i(-1)^{h+1} L\over 2^d \pi^{d+1\over 2}}
  {hx^{h-1}\over \Gamma(h+1)}
  \Gamma\left( h+{1-d\over 2}\right)
   {1\over \left(x(1-x)\left(-k_1^2 
 +m_{f+\ell}^2 \right)
  \right)^{h+{1-d\over 2}}} .
  \label{sp0}
\eea

In Eq.~(\ref{p4b}), 
the first term is
\bea
 &&  -i (p^2)^2
     \sum_{\ell =-\infty}^\infty
    \int {d^4k_1\over (2\pi)^4}
    \, {1\over k_1^2 -m_\ell^2}
  \left(
    \sum_{s=-\infty}^\infty
      \int {d^4k_2\over (2\pi)^4} \,
        {1\over k_2^2 -m_s^2}
        {1\over ((k_1+k_2)^2 -m_{f+\ell+s}^2)^3} 
      \right)
\nonumber
\\
  &\!\!\! \stackrel{s_p=0}{=}\!\!\!&
  (p^2)^2 \sum_{\ell=-\infty}^\infty
    \int{d^4k_1 \over (2\pi)^4}
    \, {1\over k_1^2 - m_\ell^2}
  \int_0^1 dx {L\over 2^d \pi^{d+1\over 2}}
    {3x^2 \over \Gamma(4)}
  {\Gamma\left({3\over 2}\right)\over \left(x(1-x)
  \left(-k_1^2 +m_{f+\ell}^2 \right)
 \right)^{3\over 2}} 
 ,
      \label{p4-1}
\eea
where $\Gamma({3\over 2})={\sqrt{\pi}\over 2}$.
With an identity
\bea
  {1\over A^\alpha B^\beta}
   =\int_0^1 dw \,
   {w^{\alpha-1} (1-w)^{\beta-1}
   \over \left[wA +(1-w)B\right]^{\alpha+\beta}} \,
   {\Gamma(\alpha+\beta)
    \over \Gamma(\alpha)\Gamma(\beta)} ,
\eea
a pair of denominators are combined as
\bea
  && \sum_{\ell=-\infty}^\infty
     \int {d^4k_{1E}\over (2\pi)^4} \,
     {1\over k_{1E}^2 +{\ell^2 \pi^2 \over L^2}}
     {1\over \left(
      k_{1E}^2 +{(f+\ell)^2 \pi^2\over L^2}
       \right)^{3\over 2}} 
\nonumber
\\
  &\!\!\!=\!\!\!&
   \sum_{\ell=-\infty}^\infty
   \int {d^4k_{1E}\over (2\pi)^4}
   \int_0^1 dw \,
   {w^{1\over 2}\over
   \left[k_{1E}^2
    +{(\ell+wf)^2\pi^2 \over L^2}
    +{w(1-w)f^2 \pi^2 \over L^2}\right]^{5\over 2}}
    \,
   {\Gamma\left({5\over 2}\right)
   \over \Gamma(1)
    \Gamma\left({3\over 2}\right)} 
    \nonumber
\\
  &\!\!\!=\!\!\!&
   \sum_{\ell=-\infty}^\infty
   \int {d^4k_{1E}\over (2\pi)^4}
   \int_0^1 dw \,
   w^{1\over 2}
 {1\over\Gamma(1)\Gamma\left({3\over 2}\right)}
   \int_0^\infty db \, b^{3\over 2}
   e^{-b\left(k_{1E}^2
    +{(\ell+wf)^2\pi^2 \over L^2}
    +{w(1-w)f^2 \pi^2 \over L^2}\right)}
\eea 
Eq.~(\ref{p4-1}), i.e., 
the first term in Eq.~(\ref{p4b}) is
\bea
 &&
  -i(p^2)^2 \sum_{\ell_p=-\infty}^\infty
   \int_0^1 dx \,
  {L\over 2^5\pi^2}
   {x^2\over 2} {1\over (x(1-x))^{3\over 2}}
   {L\over 2^{d-1} \pi^{1+{d\over 2}}}
\nonumber
\\
  && \times
   \int_0^1 dw
    \, \sqrt{w}
     \int_0^\infty db \, b^{1-{d\over 2}}
      \exp\left[-{\ell_p^2 L^2 \over b}
       +2\pi i\ell_p wf
   -{w(1-w)f^2 \pi^2\over L^2}b
  \right] .
  \label{firstapr}
\eea
This term seems to be singular near $x=1$.
After the three terms in Eq.~(\ref{p4b})
are added up,
it will be seen that the sum has no
singularity about $x$.

Another equation for the integrals 
included in Eq.~(\ref{p4b}) is
\bea
  && \sum_{s=-\infty}^\infty
   \int {d^4 k_2 \over (2\pi)^4}
    \, {1\over k_2^2 -m_s^2}
    {((k_1+k_2)^2)^u
    \over 
     ((k_1+k_2)^2 -m_{f+\ell+s}^2)^h} 
\nonumber
\\
  &\!\!\!=\!\!\!&
  \sum_{s=-\infty}^\infty
   \int_0^1 dx
    \int {d^4\ell\over (2\pi)^4}
    \,
    {hx^{h-1}\over \left[\ell^2 -\Delta\right]^{h+1}}
    ((\ell+(1-x)k_1)^2)^u 
  , \label{equ}
\eea
where $u=1$ or $u=2$.

For $u=1$, the equation (\ref{equ}) is
\bea
  &&
  \sum_{s=-\infty}^\infty
  \int_0^1 dx \int {d^4\ell_E\over (2\pi)^4}
  \, 
   {i(-1)^{h+1} hx^{h-1}\over \Gamma(h+1)}
   \int_0^\infty dt \, t^h e^{-(\ell_E^2 +\Delta)t}
   (-\ell_E^2+(1-x)^2k_1^2) 
\nonumber
\\
  &\!\!\!=\!\!\!&
   \sum_{s=-\infty}^\infty
     \int_0^1 dx        
        {i(-1)^{h+1} hx^{h-1}\over 
 (2\pi)^d \Gamma(h+1)}
        \int_0^\infty dt \, t^h e^{-\Delta t}  
  \sqrt{\pi\over t}^d
  \left(-{d\over 2t} 
  +(1-x)^2 k_1^2 \right) .
  \label{u1}
\eea
Here
\bea
  && \int_{-\infty}^\infty
     d^d \ell_E
        \, e^{-\ell_E^2 t} 
     = \sqrt{\pi\over t}^d ,
 \qquad
  \int_{-\infty}^\infty
     d^d \ell_E \, \ell_E^2 e^{-\ell_E^2 t}
  = {d\over 2t}\sqrt{\pi\over t}^d ,
\\
 && 
 \int_{-\infty}^\infty
     d^d \ell_E \, (\ell_E^2)^2 e^{-\ell_E^2 t}
    =
     {d\over 2t^2}\left({d\over 2}+1\right)
   \sqrt{\pi\over t}^d   .
\eea
For $s_p=0$ in Poisson's summation, 
Eq.~(\ref{u1}) is  
\bea
  && \int_0^1 dx 
  \, 
  {i(-1)^{h+1}L\over 2^d \pi^{d+1\over 2}}
   {hx^{h-1}\over \Gamma(h+1)}
   \Gamma\left(h-{d+1\over 2}\right)
    {d\over 2}
      {-1\over \left(x(1-x)
      \left(-k_1^2 +m_{f+\ell}^2
   \right)\right)^{h-{d+1\over 2}}}
\nonumber
\\
 && +\int_0^1 dx \,
   {i(-1)^{h+1}L\over 2^d \pi^{d+1\over 2}}
    {hx^{h-1}\over \Gamma(h+1)}
    \Gamma\left(h+{1-d\over 2}\right)
    {(1-x)^2 k_1^2 \over 
    \left(
    x(1-x) \left(-k_1^2 +
     m_{f+\ell}^2 \right)
  \right)^{h+{1-d\over 2}}} .
\nonumber
\\
  &&
\eea   
For $u=2$, Eq.~(\ref{equ}) is
\bea
 && \sum_{s=-\infty}^\infty
  \int_0^1 dx
 \int {d^4\ell_E\over (2\pi)^4} \,
 {i(-1)^{h+1} hx^{h-1}\over \Gamma(h+1)}
  \int_0^\infty dt \, t^h e^{-(\ell_E^2 +\Delta)t}
\nonumber
\\
  &&\times
  \left((\ell_E^2)^2 
 -3(1-x)^2 \ell_E^2 k_1^2
  +(1-x)^4 (k_1^2)^2 \right) .
\eea
For $s_p=0$ for Poisson's summation, this equation is
\bea
   && \int_0^1 dx \, 
   {i(-1)^{h+1} L\over 2^d \pi^{d+1\over 2}}
  {hx^{h-1}\over \Gamma(h+1)}
  \left[
  {\Gamma\left(h-{d+3\over 2}\right)
   {d\over 2}\left({d\over 2}+1\right)
  \over \left(
   x(1-x) \left(-k_1^2 +m_{f+\ell}^2
  \right) \right)^{h-{d+3\over 2}}}
   \right.
\nonumber
\\
 &&\left.
 + 
  {\Gamma\left(h-{d+1\over 2}\right)
   {d\over 2}
  (-3(1-x)^2 k_1^2) \over \left(
   x(1-x) \left(-k_1^2 +m_{f+\ell}^2
  \right) \right)^{h-{d+1\over 2}}}
  + 
  {\Gamma\left(h+{1-d\over 2}\right)
  (1-x)^4 (k_1^2)^2 \over \left(
   x(1-x) \left(-k_1^2 +m_{f+\ell}^2
  \right) \right)^{h+{1-d\over 2}}}\right] .
\eea
From these equations, the second and third terms
in Eq.~(\ref{p4b}) are derived as follows.

In Eq.~(\ref{p4b}), the second term
is
\bea
  3i (p^2)^2
   \sum_{\ell=-\infty}^\infty
     \int {d^4k_1\over (2\pi)^4}
      \, {1\over k_1^2 -m_\ell^2}
     \left(
      \sum_{s=-\infty}^\infty 
      \int {d^4k_2 \over (2\pi)^4}
      \, {1\over k_2^2 -m_s^2}
      {(k_1+k_2)^2 \over
      ((k_1+k_2)^2 - m_{f+\ell+s}^2)^4}\right)
      .
     \label{secondap}
\eea
For $s_p=0$ for Poisson's summation,
Eq.~(\ref{secondap}) is
\bea
  && 3(p^2)^2 \sum_{\ell=-\infty}^\infty
   \int {d^4k_1\over (2\pi)^4}
    \, {1\over k_1^2 - m_\ell^2}
    \int_0^1 dx \,
    {L\over 2^4\pi^{5\over 2}}
    {x^3\over 6} 
\nonumber
\\
 && \times \left(
  {\Gamma\left({3\over 2}\right)
    (-2)\over (x(1-x))^{3\over 2}}
    {1\over \left(
      -k_1^2 + m_{f+\ell}^2 \right)^{3\over 2}}
   +{\Gamma\left({5\over 2}\right)
     (1-x)^2 \over (x(1-x))^{5\over 2}}
     {k_1^2 \over 
     \left(-k_1^2 +
      m_{f+\ell}^2\right)^{5\over 2}} \right) 
\nonumber
\\
 &\!\!\!=\!\!\!&3 (p^2)^2 \int_0^1 dx \,
   {L\over 2^4 \pi^{5\over 2}}{x^3\over 6}
   {iL\over 2^{d-1} \pi^{1+{d\over 2}}}
\nonumber
\\
  && \times
  \left(
   {\Gamma\left({3\over 2}\right) 
    \cdot 2\over (x(1-x))^{3\over 2}}    
    \int_0^1 dw \, \sqrt{w}
     \int_0^\infty db \, b^{1-{d\over 2}}
     e^{-bw(1-w) {f^2\pi^2 \over L^2}}
   \right.
\nonumber
\\
  && \left.
    + 
 {\Gamma\left({5\over 2}\right) 
    (1-x)^2\over (x(1-x))^{5\over 2}}
    \int_0^1 dw \, w^{3\over 2} {d\over 3}
     \int_0^\infty db \, b^{1-{d\over 2}}
     e^{-bw(1-w) {f^2\pi^2 \over L^2}}
  \right),
  \label{secondapr}
\eea
where $\ell_p =0$ for Poisson's summation has been taken.

In Eq.~(\ref{p4b}), the third term is
\bea
   -2i(p^2)^2 
    \sum_{\ell=-\infty}^\infty
       \int {d^4 k_1 \over (2\pi)^4}
       \, {1\over k_1^2 - m_\ell^2}
       \left(
     \sum_{s=-\infty}^\infty
     \int{d^4k_2\over (2\pi)^4}
     \, {1\over k_2^2 -m_s^2}
     {((k_1+k_2)^2)^2
     \over ((k_1+k_2)^2 -
      m_{f+\ell+s}^2)^5} \right) .
  \label{thirdap}
\eea
For $s_p=0$ for Poisson's summation,
Eq.~(\ref{thirdap}) is
\bea
  && 2(p^2)^2 \sum_{\ell=-\infty}^\infty
    \int {d^4 k_1 \over (2\pi)^4}
    \, {1\over k_1^2-{\ell^2 \pi^2 \over L^2}}
    \int_0^1 dx \,{L\over 2^4\pi^{5\over 2}}
    {x^4\over 24}  
 \left(
  {\Gamma\left({3\over 2}\right)
    \cdot 6\over (x(1-x))^{3\over 2}}
     {1\over \left( -k_1^2 +
      m_{f+\ell}^2 \right)^{3\over 2}}
   \right.
\nonumber
\\
  && \left.
  +{\Gamma\left({5\over 2}\right)
   (-6(1-x)^2)\over (x(1-x))^{5\over 2}}
     {k_1^2\over 
     \left(-k_1^2 +m_{f+\ell}^2 \right)^{5\over 2}}
  +{\Gamma\left({7\over 2}\right)
    (1-x)^4\over (x(1-x))^{7\over 2}}
     {(k_1^2)^2\over
     \left(-k_1^2 +
     m_{f+\ell}^2 \right)^{7\over 2}}
     \right)
\nonumber
\\
  &\!\!\!=\!\!\!&
  2(p^2)^2 
    \int_0^1 dx \,{L\over 2^4\pi^{5\over 2}}
    {x^4\over 24} 
        {-iL\over 2^{d-1}\pi^{1+{d\over 2}}}
\nonumber
\\
  && \times
 \left(
  {\Gamma\left({3\over 2}\right)
    \cdot 6\over (x(1-x))^{3\over 2}}
      \int_0^1 dw \, \sqrt{w}
       \int_0^\infty db\, b^{1-{d\over 2}}
        e^{-bw(1-w){f^2 \pi^2 \over L^2}}
   \right.
\nonumber
\\
  && +{\Gamma\left({5\over 2}\right)
   \cdot 6(1-x)^2\over (x(1-x))^{5\over 2}}
     \int_0^1 dw \, w^{3\over 2}{d\over 3}
       \int_0^\infty db\, b^{1-{d\over 2}}
        e^{-bw(1-w){f^2 \pi^2 \over L^2}}
\nonumber
\\
  && \left.
  +{\Gamma\left({7\over 2}\right)
    (1-x)^4\over (x(1-x))^{7\over 2}}
     \int_0^1 dw \, w^{5\over 2} {d(d+2)\over 15}
       \int_0^\infty db\, b^{1-{d\over 2}}
        e^{-bw(1-w){f^2 \pi^2 \over L^2}}
     \right) ,
     \label{thirdapr}
\eea
where $l_p=0$ has been taken.

The sum of Eqs.~(\ref{firstapr}),
(\ref{secondapr}) and (\ref{thirdapr}), i.e.,
the two-loop contribution (\ref{p4b}) is
\bea
  i(p^2)^2 
  {L^2\over 2^9\pi^5} \int_0^1 dx \,
  \sqrt{x(1-x)} 
    \int_0^1 dw \,
    (w^{1\over 2} -2 w^{3\over 2} + w^{5\over 2})
 \int_0^\infty db\, 
  b^{1-{d\over 2}} e^{-bw(1-w){f^2\pi^2 \over L^2}} 
 . 
  \label{total2ap}
\eea
The integral with respect to $b$ is found to 
have logarithmic 
divergence from
\bea
  \int_0^\infty db \, b^{1-{d\over 2}}
     e^{-w(1-x){f^2\pi^2 \over L^2}b}
    =
  {\Gamma\left(2-{d\over 2}\right) 
 \over \left({w(1-w)f^2\pi^2 \over L^2}
  \right)^{2-{d\over 2}}} 
  =
   \int {d^d q_E\over (2\pi)^d}
     {(4\pi)^{d\over 2}\over
       \left(q_E^2 + {w(1-w)f^2 \pi^2 \over L^2}\right)^2
         } ,
\eea
and a cutoff regularization
\bea
 \int {d^4 q_E\over (2\pi)^4}
     {(4\pi)^2\over
       \left(q_E^2 + {w(1-w)f^2 \pi^2 \over L^2}\right)^2
         } 
 = \log \left({\Lambda^2 L^2 \over 
      w(1-w) f^2 \pi^2}\right)
      +1 .
\eea
Thus the divergence of Eq.~(\ref{total2ap}) is
obtained as Eq.~(\ref{2loopr}) with
\bea 
  \int_0^1 dx \sqrt{x(1-x)} 
 = {\pi \over 8} ,
\qquad
 \int_0^1 dw\, 
  (w^{1\over 2} -2 w^{3\over 2} + w^{5\over 2}) 
  = {16\over 105} .
\eea

\section{Expansion of products of $\delta$
\label{ap:d}}

In this section,
we give an equation for
expansion of the following product
of $\delta_{|Q||R|}=\delta_{QR} +\delta_{Q,-R}
-\delta_{Q0}\delta_{R0}$:
\bea
 &&
     \delta_{|f||n|} \,
     \delta_{|g||a|} \,
     \delta_{|\ell||b|} \,
     \delta_{|s||c|} \,
     \delta_{|n+\ell+s||a+b+c|}
\nonumber
\\
  &\!\!\!=\!\!\!&
     (\delta_{fn}+\delta_{f,-n}
     -\delta_{f0}\delta_{n0}) 
     (\delta_{ga}+ \delta_{g,-a}
    -\delta_{g0}\delta_{a0}) 
     (\delta_{\ell b} + \delta_{\ell, -b}
    -\delta_{\ell 0}\delta_{b0}) 
\nonumber
\\
  && \times
     (\delta_{sc}+\delta_{s,-c}
    -\delta_{s0}\delta_{c0}) 
     (\delta_{n+\ell+s,a+b+c}
    +\delta_{n+\ell+s,-a-b-c}
    -\delta_{n+\ell+s,0}\delta_{a+b+c,0}) .
    \label{prod}
\eea
Here we introduce
symbols $+,-,0$ for $\delta_{Q,R},
\delta_{Q,-R},-\delta_{Q0}\delta_{R0}$ in 
$\delta_{|Q||R|}$, respectively.
Components of
$\delta_{fn}, \delta_{ga},\delta_{\ell b},\delta_{sc}$
are specified with four symbols.
For example, the component ${}^+_{+++}$
expresses 
$\delta_{fn}\times 
     \delta_{ga}
     \delta_{\ell b} 
     \delta_{sc} 
   (\delta_{n+\ell+s,a+b+c}
    +\delta_{n+\ell+s,-a-b-c}
    -\delta_{n+\ell+s,0}\delta_{a+b+c,0})$.
After $\sum_n \sum_a \sum_b \sum_c$ are summed,
Eq.~(\ref{prod}) is
\bea
&& (\delta_{f, g} + \delta_{f + g + 2 (l + s), 0} - \delta_{f + l + s, 0} \delta_{f, g}) {}^+_{+++}
  + (\delta_{f + g + 2 (l + s), 0} + \delta_{f,  g} - \delta_{f + l + s, 0} \delta_{f, g})  {}^+_{---}
\nonumber
\\ &&   
  +(\delta_{f + g, 2 (l + s)} + \delta_{f, g} - \delta_{-f +l + s, 0} \delta_{f,  g})    {}^-_{+--}
  +(\delta_{f, g} + \delta_{f + g, 2 (l + s)} - \delta_{-f +l + s, 0} \delta_{f,  g})     {}^-_{-++}
\nonumber
\\ &&   
\\ &&     
  +(\delta_{f + 2 s, g} + \delta_{f + g + 2 l,   0} - \delta_{f + l + s, 0} \delta_{f + 2 s, g}) {}^+_{++-}
  +(\delta_{f + 2 l, g} + \delta_{f + g + 2 s,   0} - \delta_{f + l + s, 0} \delta_{f + 2 l, g})  {}^+_{+-+} \nonumber
\\ &&   
  +(\delta_{f + g + 2 s, 0} + \delta_{f + 2 l,  g} - \delta_{f + l + s, 0} \delta_{f + 2 l, g})  {}^+_{-+-}
  + (\delta_{f + g + 2 l, 0} + \delta_{f + 2 s,  g} - \delta_{f + l + s, 0} \delta_{f + 2 s, g}) {}^+_{--+}
\nonumber
\\ &&
  +(\delta_{f + g, 2 s} + \delta_{f,  g + 2 l} - \delta_{-f +l + s, 0} \delta_{f,  g + 2 l})   {}^-_{++-}
  +(\delta_{f + g, 2 l} + \delta_{f,   g + 2 s} - \delta_{-f +l + s, 0} \delta_{f,  g + 2 s})  {}^-_{+-+} 
\nonumber
\\ &&
  +(\delta_{f, g + 2 s} + \delta_{f + g,   2 l} - \delta_{-f +l + s, 0} \delta_{f,  g + 2 s})  {}^-_{-+-}
 +(\delta_{f, g + 2 l} + \delta_{f + g,   2 s} - \delta_{-f +l + s, 0} \delta_{f, g + 2 l})   {}^-_{--+}
\nonumber
\\ &&  
\\ &&  
  - \delta_{s, 0} (\delta_{f, g} + \delta_{f + g + 2 l,   0} - \delta_{f + l, 0} \delta_{f, g})  {}^+_{++0}
  -\delta_{l,  0} (\delta_{f, g} + \delta_{f + g + 2 s,   0} - \delta_{f + s, 0} \delta_{f,  g}) {}^+_{+0 +}
\nonumber
\\ &&
  -  \delta_{s,  0} (\delta_{f + g + 2 l, 0} + \delta_{f,  g} - \delta_{f + l, 0} \delta_{f, g})  {}^+_{--0}
  -  \delta_{l,  0} (\delta_{f + g + 2 s, 0} + \delta_{f,  g} - \delta_{f + s, 0} \delta_{f, g}) {}^+_{-0 -} 
\nonumber
\\ &&
   -\delta_{s,  0} (\delta_{f + g, 2 l} + \delta_{f,  g} - \delta_{-f + l, 0} \delta_{f,  g}) {}^-_{+-0}
   -\delta_{l,  0} (\delta_{f + g, 2 s} + \delta_{f,  g} - \delta_{-f + s, 0} \delta_{f,   g}) {}^-_{+0 -}
\nonumber
\\ &&
  - \delta_{s,  0} (\delta_{f, g} + \delta_{f + g,   2 l} - \delta_{-f + l, 0} \delta_{f,   g})  {}^-_{-+0}
  - \delta_{l,  0} (\delta_{f, g} + \delta_{f + g,   2 s} - \delta_{-f + s, 0} \delta_{f,  g})  {}^-_{-0 +}
\nonumber
\\ &&  
\\ &&  
  +(\delta_{f + 2 (l + s), g} + \delta_{f + g,   0} - \delta_{f + l + s, 0} \delta_{f + g, 0})  {}^+_{+--}
  +(\delta_{f + g, 0} + \delta_{f + 2 (l + s),   g} - \delta_{f + l + s, 0} \delta_{f + g, 0})   {}^+_{-++}
\nonumber
\\ &&
  +(\delta_{f + g, 0} + \delta_{f,  g + 2 (l + s)} - \delta_{-f +l + s, 0} \delta_{f + g,   0})  {}^-_{+++}
\nonumber
\\
 &&
  +(\delta_{f, g + 2 (l + s)} + \delta_{f + g,  0} - \delta_{-f +l + s, 0} \delta_{f + g,   0})  {}^-_{---}
\nonumber
\\ &&  
\\ &&  
  -\delta_{s, 0} (\delta_{f + 2 l, g} + \delta_{f + g,  0} - \delta_{f + l, 0} \delta_{f + g, 0}) {}^+_{+-0}
  -\delta_{l, 0} (\delta_{f + 2 s, g} + \delta_{f + g, 0} - \delta_{f + s, 0} \delta_{f + g, 0}) {}^+_{+0 -}
\nonumber
\\ &&
  -\delta_{s, 0} (\delta_{f + g, 0} + \delta_{f + 2 l,  g} - \delta_{f + l, 0} \delta_{f + g, 0}) {}^+_{-+0}
  -\delta_{l, 0} (\delta_{f + g, 0} + \delta_{f + 2 s,  g} - \delta_{f + s, 0} \delta_{f + g, 0}) {}^+_{-0 +}
\nonumber
\\ &&
  -\delta_{s, 0} (\delta_{f + g, 0} + \delta_{f,  g + 2 l} - \delta_{-f + l, 0} \delta_{f + g, 0}) {}^-_{++0}
  -\delta_{l, 0} (\delta_{f + g, 0} + \delta_{f,  g + 2 s} - \delta_{-f + s, 0} \delta_{f + g, 0}) {}^-_{+0 +}
\nonumber
\\ &&
 -\delta_{s, 0} (\delta_{f, g + 2 l} + \delta_{f + g, 0} - \delta_{-f + l, 0} \delta_{f + g , 0}) {}^-_{--0}
 -\delta_{l, 0} (\delta_{f, g + 2 s} + \delta_{f + g,0} - \delta_{-f + s, 0} \delta_{f + g, 0}) {}^-_{-0 -}   
\nonumber
\\ &&
\\ &&    
 + \delta_{l,0} \delta_{s,0} (\delta_{f, g} + \delta_{f + g, 0} - \delta_{f, 0} \delta_{g, 0}) {}^+_{+00}    
 +\delta_{l, 0} \delta_{s,0} (\delta_{f + g, 0} + \delta_{f, g} - \delta_{f, 0} \delta_{g, 0}) {}^+_{-00}
\nonumber
\\ &&
 +\delta_{l, 0} \delta_{s,0} (\delta_{f + g, 0} + \delta_{f, g} - \delta_{f, 0} \delta_{g, 0}) {}^-_{+00}
   +\delta_{l, 0} \delta_{s, 0} (\delta_{f, g} + \delta_{f + g, 0} - \delta_{f, 0} \delta_{g , 0}) {}^-_{-00} 
\nonumber
\\ &&             
\\ &&                                                    
  -\delta_{g,  0} (\delta_{f, 0} + \delta_{f + 2 (l + s),  0} - \delta_{f, 0} \delta_{l + s, 0})  {}^+_{0++}
  -\delta_{g,  0}(\delta_{f + 2 (l + s), 0} + \delta_{f, 0} - \delta_{f, 0} \delta_{l + s, 0})  {}^+_{0--}
\nonumber
\\ &&
- \delta_{g,  0} (\delta_{f, 0} + \delta_{f, 2 (l + s)} - \delta_{f, 0} \delta_{ l + s,  0}) {}^-_{0++}
  -\delta_{g,  0} (\delta_{f, 2 (l + s)} + \delta_{f,  0} - \delta_{f, 0} \delta_{l + s, 0})  {}^-_{0--}
\nonumber
\\ &&  
\\ &&  
  -\delta_{g,  0} (\delta_{f + 2 s, 0} + \delta_{f + 2 l, 0} - \delta_{f + 2 l, 0} \delta_{l, s}) {}^+_{0 + -}
  -\delta_{g,  0} (\delta_{f + 2 l, 0} + \delta_{f + 2 s, 0} - \delta_{f + 2 s, 0} \delta_{l, s}) {}^+_{0 - +}
\nonumber
\\ &&
  - \delta_{g, 0} (\delta_{f, 2 s} + \delta_{f,  2 l} - \delta_{f, 2 l} \delta_{l,  s})   {}^-_{0 + -}
  -\delta_{g,  0} (\delta_{f, 2 l} + \delta_{f,  2 s} - \delta_{f, 2 l} \delta_{l,  s})   {}^-_{0 - +}
\nonumber
\\ &&  
\\ &&  
 +\delta_{g, 0} \delta_{s,  0} (\delta_{f, 0} + \delta_{f + 2 l,0} - \delta_{f, 0} \delta_{l, 0}) {}^+_{0 + 0}
  +\delta_{g, 0} \delta_{s,0} (\delta_{f + 2 l, 0} + \delta_{f, 0} - \delta_{f, 0} \delta_{l, 0}) {}^+_{0 - 0}
\nonumber
\\ &&
+\delta_{g,0} \delta_{l,0} (\delta_{f, 0} + \delta_{f +2 s, 0} - \delta_{f,0} \delta_{s,0}) {}^+_{00 +}        
 +\delta_{g, 0} \delta_{l,0} (\delta_{f + 2 s, 0} + \delta_{f, 0} - \delta_{f, 0} \delta_{s, 0}) {}^+_{00 -}
\nonumber
\\ &&
 +\delta_{g, 0} \delta_{s, 0} (\delta_{f, 0} + \delta_{f, 2 l} - \delta_{f, 0} \delta_{l, 0})  {}^-_{0+0}
 + \delta_{g, 0} \delta_{s,  0} (\delta_{f, 2 l} + \delta_{f, 0} - \delta_{f, 0} \delta_{l, 0}) {}^-_{0- 0}
\nonumber
\\ &&
  +\delta_{g,0} \delta_{l,0} (\delta_{f, 0} +\delta_{f, 2 s} -\delta_{f, 0} \delta_{s, 0}) {}^-_{00 +}        
 +\delta_{g, 0} \delta_{l, 0} (\delta_{f, 2 s} + \delta_{f, 0} - \delta_{f, 0} \delta_{s, 0}) {}^-_{00 -}
\nonumber
\\ &&  
\\ &&  
  -(\delta_{g, 0} \delta_{l, 0} \delta_{s, 0} \delta_{f, 0})   {}^+_{000}                                  
  - (\delta_{g, 0} \delta_{l, 0} \delta_{s, 0} \delta_{f, 0})      {}^-_{000}                                  
\nonumber
\\ &&
  -(\delta_{f, 0} \delta_{l, 0} \delta_{s, 0} \delta_{0,  g})   {}^0_{+00}                           
  - (\delta_{f, 0} \delta_{l, 0} \delta_{s, 0} \delta_{0,  g})  {}^0_{-00}
\nonumber
\\ &&
  + (\delta_{f, 0} \delta_{g, 0} \delta_{l, 0} \delta_{s, 0})   {}^0_{000}   
\nonumber
\\ &&
\\ &&    
  -\delta_{f,  0} (\delta_{0, g} + \delta_{g + 2 (l + s),   0} - \delta_{l + s, 0} \delta_{g, 0}) {}^0_{+++}
  -\delta_{f,  0} (\delta_{2 (l + s), g} + \delta_{0, g} - \delta_{l + s, 0} \delta_{g, 0}) {}^0_{+--}
\nonumber
\\ &&
  - \delta_{f, 0} (\delta_{0, g} + \delta_{2 (l + s),  g} - \delta_{l + s, 0} \delta_{g,  0})  {}^0_{-++}
  - \delta_{f,  0} (\delta_{g + 2 (l + s), 0} + \delta_{0, g} - \delta_{l + s, 0} \delta_{g, 0}) {}^0_{---}
\nonumber
\\ &&  
\\ &&  
  -\delta_{f,  0} (\delta_{g, 2 s} + \delta_{g + 2 l,0} - \delta_{l + s, 0} \delta_{g + l - s,0}) {}^0_{++-}
  -\delta_{f, 0} (\delta_{2 l, g} + \delta_{g + 2 s, 0} - \delta_{l + s, 0} \delta_{g, 2 l}) {}^0_{+-+} 
\nonumber
\\ &&
  -\delta_{f,0} (\delta_{g + 2 s, 0} + \delta_{2 l, g} - \delta_{l + s, 0} \delta_{-g + l - s,0})  {}^0_{-+-}
 -\delta_{f,0} (\delta_{g + 2 l, 0} + \delta_{2 s, g} - \delta_{l + s, 0} \delta_{-g - l + s, 0}) {}^0_{--+}
\nonumber
\\ &&
\\ &&
 +\delta_{f, 0} \delta_{s, 0} (\delta_{0, g} + \delta_{g + 2 l,  0} - \delta_{l, 0} \delta_{ g, 0}) {}^0_{++0}
 +\delta_{f, 0} \delta_{s, 0} (\delta_{2 l, g} + \delta_{0,g} - \delta_{l, 0} \delta_{g, 0}) {}^0_{+-0}
\nonumber
\\ &&
 +\delta_{f, 0} \delta_{l, 0} (\delta_{0, g} + \delta_{g + 2 s,  0} - \delta_{s, 0} \delta_{g, 0}) {}^0_{+0 +}
 +\delta_{f, 0} \delta_{l, 0} (\delta_{2 s, g} + \delta_{0, g} - \delta_{s, 0} \delta_{g, 0})  {}^0_{+0 -}
\nonumber
\\ &&
 + \delta_{f, 0} \delta_{s, 0} (\delta_{0, g} + \delta_{2 l, g} - \delta_{l, 0} \delta_{g, 0}) {}^0_{-+0}
 + \delta_{f, 0} \delta_{s, 0} (\delta_{g + 2 l, 0} + \delta_{0, g} - \delta_{l, 0} \delta_{g, 0}) {}^0_{--0}
\nonumber
\\ &&
 + \delta_{f, 0} \delta_{l, 0} (\delta_{0, g} + \delta_{2 s, g} - \delta_{s, 0} \delta_{g, 0}) {}^0_{-0 +}
 + \delta_{f, 0} \delta_{l, 0} (\delta_{g + 2 s, 0} + \delta_{0, g} - \delta_{s, 0} \delta_{g, 0}) {}^0_{-0 -}
\nonumber
\\ &&
\\ &&                                         
+ (\delta_{f, 0} \delta_{g,  0})     {}^0_{0++}
 +(\delta_{f, 0} \delta_{g,  0})  {}^0_{0--}
\nonumber
\\ &&
\\ &&
+ \delta_{f, 0} \delta_{g,  0} (\delta_{s, 0} + \delta_{l,0} - \delta_{l, 0} \delta_{s,  0}) {}^0_{0 + -}
 +\delta_{f, 0} \delta_{g,  0} (\delta_{l, 0} + \delta_{s, 0} - \delta_{l, 0} \delta_{s,  0}) {}^0_{0 - +}
\nonumber
\\ &&
\\ &&
 -(\delta_{f, 0} \delta_{g, 0} \delta_{s,  0}) {}^0_{0 + 0}
 - (\delta_{f, 0} \delta_{g, 0} \delta_{s,  0}) {}^0_{0- 0}
 - (\delta_{f, 0} \delta_{g,0} \delta_{l,0}) {}^0_{00+} 
 - (\delta_{f, 0} \delta_{g, 0} \delta_{l, 0}) {}^0_{00 -}
 .
\nonumber
\\
 &&  
  \label{basic}
\eea
where symbols of classification have been
shown with superscripts and subscripts
and equation numbers have been inserted at the points
for grouping $3^4$ terms. 
By substituting this equation into Eq.~(\ref{44}),
the two-loop correction (\ref{2loop44}) is obtained.
 
\end{appendix}

\newpage



\end{document}